\documentclass[10pt,journal,compsoc]{IEEEtran}

\usepackage{tabularx}
    \newcolumntype{L}{>{\raggedright\arraybackslash}X}
\usepackage{comment}
\usepackage{amsmath}
\usepackage{amssymb}
\usepackage{gensymb}
\usepackage{graphicx}
\usepackage{epstopdf} 
\usepackage{color}
\usepackage{enumerate}
\usepackage{multirow}
\usepackage{cite}
\usepackage{multicol}
\usepackage{threeparttable}
\usepackage[ruled]{algorithm}
\usepackage[noend]{algorithmic}
\usepackage{subfig}
\usepackage{float}
\usepackage{mathtools}
\usepackage{array}
\newcolumntype{P}[1]{>{\centering\arraybackslash}p{#1}}
\newcolumntype{M}[1]{>{\centering\arraybackslash}m{#1}}
\usepackage{nopageno}

\usepackage[numbers]{natbib} 

\begin{document}

\title{SHARKS: Smart Hacking Approaches for RisK Scanning in Internet-of-Things and Cyber-Physical Systems based on Machine Learning}

\author{Tanujay Saha, Najwa Aaraj, Neel Ajjarapu and Niraj K. Jha~\IEEEmembership{(Fellow,~IEEE)}

\IEEEcompsocitemizethanks{\IEEEcompsocthanksitem 
T. Saha, N. Ajjarapu and N. K. Jha are with 
the Department of Electrical Engineering, Princeton University, New Jersey,
NJ, 08544 (\{tsaha,najjarapu,jha\}@princeton.edu.  N. Aaraj is with Technology Innovation Institute, UAE (najwa@tii.ae).
}
}

\IEEEtitleabstractindextext{
\begin{abstract}
Cyber-physical systems (CPS) and Internet-of-Things (IoT) devices are
increasingly being deployed across multiple functionalities, ranging
from healthcare devices and wearables to critical infrastructures, e.g.,
nuclear power plants, autonomous vehicles, smart cities, and smart
homes. These devices are inherently not secure across their comprehensive
software, hardware, and network stacks, thus presenting a large
attack surface that can be exploited by hackers. In this article,
we present an innovative technique for detecting {\em unknown} system
vulnerabilities, managing these vulnerabilities, and improving incident
response when such vulnerabilities are exploited. The novelty of this
approach lies in extracting intelligence from known real-world CPS/IoT
attacks, representing them in the form of regular expressions, and
employing machine learning (ML) techniques on this ensemble of regular
expressions to generate new attack vectors and security vulnerabilities. Our 
results show that 10 new attack vectors and 122 new vulnerability exploits can 
be successfully generated that have the potential to exploit a CPS or an 
IoT ecosystem. The ML methodology achieves an accuracy of 97.4\% and enables 
us to predict these attacks efficiently with an 87.2\% reduction in the search 
space. We demonstrate the application of our method to the hacking of the in-vehicle 
network of a connected car. To defend against the known attacks and possible 
novel exploits, we discuss a defense-in-depth mechanism for various classes 
of attacks and the classification of data targeted by such attacks. This 
defense mechanism optimizes the cost of security measures based on the 
sensitivity of the protected resource, thus incentivizing its adoption in 
real-world CPS/IoT by cybersecurity practitioners.
\end{abstract}

}

\maketitle
\IEEEdisplaynontitleabstractindextext
\IEEEpeerreviewmaketitle

\IEEEraisesectionheading{\section{Introduction}}
\label{Intro}
\IEEEPARstart{C}{yber-physical} systems (CPS) use sensors to feed data to 
computing elements that monitor and control physical systems and use actuators to elicit desired changes in the environment.  Internet-of-Things (IoT) enables diverse, uniquely identifiable, and resource-constrained devices 
(sensors, processing elements, actuators) to exchange data through the 
Internet and optimize desired processes.  CPS/IoT have a plethora of
applications, like smart cities~\cite{arasteh2016iot}, smart healthcare \cite{akmandor2018smart}, smart homes, nuclear plants, smart grids 
\cite{yun2010research}, autonomous vehicles 
\cite{datta2016integrating}, and in various other domains. With recent 
advances in CPS/IoT-facilitating technologies like machine learning (ML), 
cloud computing, and 5G communication systems 
\cite{DBLP:journals/scn/HuangCLY16}, CPS/IoT are likely to have an even
more widespread impact in the near future.

\par An unfortunate consequence of integrating multiple devices in an ecosystem is the dramatic 
increase in its attack surface. Most of the CPS/IoT devices are energy-constrained, which makes them 
unable to implement existing elaborate cryptographic protocols and primitives as well as other 
conventional security measures across the software, hardware, and network stacks 
\cite{mosenia2017comprehensive, xu2014security}. The diverse range of embedded devices in the network 
and inherent vulnerabilities in the design and implementation, coupled with an absence of standard 
cryptographic primitives and network security protocols, make CPS/IoT a favorable 
playground for malicious attackers. Although lightweight cryptographic 
protocols \cite{katagi2008lightweight,lee2014lightweight} and hardware-based 
(lightweight) authentication protocols \cite{saha2016tv,DBLP:conf/dac/SuhD07, 
7829547} mitigate some threats, most of the vulnerabilities remain 
unaddressed. Another challenge in securing CPS/IoT is the large amount 
of accessible data generated by the numerous communication channels among  
devices. Such data, in the absence of adequate cryptographic
technologies, pose a threat to the CPS/IoT device and consequently
impact user privacy, data confidentiality, and integrity. Moreover, CPS/IoT are 
vulnerable to a plethora of attacks \cite{6188257, mosenia2017comprehensive}, 
e.g., buffer overflow exploits, race conditions, XSS attacks that target known 
vulnerabilities, and new (undiscovered) vulnerabilities, the exploit of which 
is referred to as a zero-day attack. 

\par In this article, we propose an ML-based approach to systematically 
generate new exploits in a CPS/IoT framework. We call this approach SHARKS, which is an acronym for Smart Hacking Approaches for RisK Scanning. ML has already found use in 
CPS/IoT cybersecurity \cite{saha2021sharks, 8454402, 
copty2018deep}, primarily in network intrusion and anomaly detection systems 
\cite{zhang2005intrusion, brown2021gravitas}. These systems execute ML algorithms on data 
generated by network logs and communication channels. In the methodology that 
we propose, ML instead operates at both system and user 
levels to predict unknown exploits against CPS/IoT. 

\par SHARKS is based on developments along two important directions. Recognizing the need to depart from the traditional approaches to cybersecurity, we observe that the main objective of many security attacks on CPS/IoT is to modify the behavior of the end-system to cause unsafe operations. Based on this insight, we propose to model the behavior of CPS/IoT under attack, at the system and network levels, use
ML to discover a more exhaustive potential attack space, and 
then map it to a defense space. Our approach enables a preemptive analysis of vulnerabilities
across a large variety of devices by detecting new
attacks and deploying patches ahead of time.

\par We analyze an exhaustive set of real-world CPS/IoT attacks 
that have been documented and represent them as regular expressions. An ML 
algorithm is then trained with these regular expressions. The trained ML
model can predict the feasibility of a new attack. The vulnerability
exploits predicted to be highly feasible by the ML algorithm are
reported as novel exploits. This approach successfully generated 122
novel exploits and 10 unexploited attack vectors. To demonstrate the
applicability of our approach, we evaluate the trained model on the
in-vehicle network of a connected car. The model was successful in
discovering 67 vulnerability exploits in the car network.

\par The novelty of the proposed methodology lies in:

\begin{itemize}
    \item Representation of real-world CPS/IoT attacks in the form of
regular expressions and control-data flow graphs (CDFGs), where both control flow and data invariants are instrumented at the system level. 
    \item Creation of an aggregated attack directed acyclic graph (DAG) with an ensemble of such regular expressions.
    \item Use of an ML model trained with these regular expressions to 
generate novel exploits in a given CPS/IoT framework.
\end{itemize}

\par The article is organized as follows. Section~\ref{section:Related_work} 
provides a summary of the work that has been done in the application of 
ML and automation to cybersecurity. 
Section~\ref{section:Background} discusses background material. Section~\ref{section::Methodology} gives details of our methodology and the results obtained with it. Section~\ref{section:Case_study} describes the 
application of our algorithm to a connected vehicle. 
Section~\ref{section:Security} proposes a tiered-security framework, composed of defense DAGs, for protection against security vulnerabilities. 
Section~\ref{section:Conclusion} concludes the article. 

\section{Related Work}
\label{section:Related_work}
In this section, we discuss some of the major works that have been done to 
automate security for real-world threat mitigation. Many major classes of 
security vulnerabilities, like memory corruption bugs and network intrusion vulnerabilities, can be detected using automation techniques. 
The domain of cybersecurity and embedded security that has been highly influenced by the popularity of ML is intrusion detection systems (IDSs), in particular an IDS targeted 
at network-level attacks. Prior to the rapid
advancements in ML, IDSs consisted of signature-based methods and 
anomaly-based techniques to detect intrusions in the network or the host 
systems.  Proposed IDSs perform quite well but have their drawbacks. Signature-based methods require regular updates of the software and are unable to detect zero-day vulnerabilities. Anomaly-based methods can detect zero-day vulnerabilities but have a very high false alarm rate (FAR). The advent of ML alleviated some of these drawbacks and thus ML was widely adopted in IDSs. Researchers have used a wide variety of ML methodologies to tackle this problem, such as unsupervised learning~\cite{sacramento2018flowhacker}, artificial neural 
networks~\cite{cannady1998artificial,bivens2002network}, Bayesian 
networks~\cite{livadas2006using, benferhat2008naive}, clustering 
methods~\cite{hendry2008intrusion, blowers2014machine, sequeira2002admit}, 
decision trees~\cite{bilge2011exposure,kruegel2003using}, ensemble learning like random forests~\cite{bilge2012disclosure,gharibian2007comparative}, 
hidden Markov models~\cite{aarnes2006using}, and support vector machines 
(SVMs)~\cite{amiri2011mutual,li2012efficient,saha2022system}. More advanced deep learning 
based IDSs use generative adversarial networks~\cite{chenGAN_IDS} and 
autoencoders~\cite{meidan2018n}. These methods provide a reactive security 
mechanism for detecting ongoing attacks. They also require significant
computational overhead because the models need to be continuously trained on recent data and all incoming traffic must be processed by the ML model before it can be catered to by the system. Our method differs from these methods in that it provides proactive security and requires zero run-time overhead.

ML has recently been used for malware and rootkit detection on mobile
devices~\cite{7891134,bickford2011security,brown2021gravitasarxiv,saha2022thesis}. These methods analyze the application programming
interface (API) call logs to detect malicious behavior. Another ML-based malware detection method 
analyzes the hardware performance counters (HPCs) to detect malware execution 
at run-time~\cite{sayadi2018customized}. A drawback of ML systems that are trained on API call logs, 
HPCs, and network logs, is that they are only able to detect the types of malware they have been 
trained on. A novel malware with completely different behavior and signature will go undetected by 
these detectors. They would also face difficulties in detecting the same malware running on 
a different platform and operating system. Our method, on the other hand, trains on application-independent representations of  
system-level attacks, which makes it platform-independent and equips it with the intelligence to detect a 
much broader class of security breaches.

Attack graphs have been widely used for analyzing the security of systems and networks~\cite{attackgraph_shandilya,saha2022machine}. Generating attack graphs has been a longstanding challenge due to the state explosion problem. Various automation techniques, like model checking~\cite{1021806}, rule-based artificial intelligence, and 
ML~\cite{aksu2018automated,saha2021machinearxiv}, have been used to tackle this challenge. Analysis of the attack graphs is also a challenge due to the enormous size and complexity of the graphs. Graph-based neural 
networks~\cite{attack_graphGNN} and reinforcement 
learning~\cite{8455909} have been used to analyze attack graphs to detect 
vulnerabilities. This article uses attack graphs at a higher granularity to
detect vulnerabilities and the exploits thereof across the entire hardware, software, and network stacks of CPS/IoT. In previous works, system-specific attack graphs have been used for vulnerability analysis. In this article, we propose a generalized attack graph that can be applied to detect vulnerabilities (and exploits thereof) in any CPS/IoT. We buttress this claim by applying our approach to detect vulnerabilities in the in-vehicle network of a connected car.

Memory corruption bugs have been a longstanding vulnerability in computer 
systems. A detailed analysis of this problem is provided 
in~\cite{szekeres2013sok}. Automation attempts have also been made for 
detecting such bugs. In~\cite{gao2018comprehensive}, static analysis is used to detect memory corruption vulnerabilities.

The discovery of hardware vulnerabilities like Spectre~\cite{kocher2019spectre} and Meltdown~\cite{lipp2018meltdown} in 
2018 opened the door to new classes of side-channel attacks on device
microarchitecture. An automated side-channel vulnerability detection 
technique for microarchitectures is proposed in~\cite{trippel2018checkmate}. Our method aims to achieve a similar goal, but across the entire hardware, software, and network stacks. 

\section{Background}
\label{section:Background}
We model existing CPS/IoT attacks as regular expressions 
and CDFGs. We train a popular ML model, namely SVM, with these CDFGs to 
predict new vulnerability exploits. This section provides 
an introduction to regular expressions, CDFGs, and SVM models 
that is required for ease of comprehending the rest of the article.

\subsection{Regular Expressions}
A regular expression is used to denote a set of string patterns. We 
use regular expressions to represent known CPS/IoT attacks
in a compact and coherent manner. 

\par The set of all possible characters permissible in a regular expression 
is referred to as its alphabet $\Sigma$. The basic operations permitted in 
regular expressions are \cite{kohavi2009switching}:

\begin{itemize}
    \item[$\bullet$]\textbf{Set union}: This represents the set union of two regular 
expressions. For example, if expression $A$ denotes $\{xy, z\}$ and $B$ 
denotes $\{xy, r, pq\}$, then expression $A + B$ denotes $\{xy, z, r, pq\}$.
    \item[$\bullet$]\textbf{Concatenation}: This operation represents the 
set of strings obtained by attaching any string in the first
expression with any string in the second expression. For example, if 
$A = \{xy, z\}$ and $B = \{r, pq\}$, then, $AB = \{xyr, xypq, zr, zpq\}$.
    \item[$\bullet$]\textbf{Kleene star}: $A^{*}$ denotes the set of strings 
obtained by concatenating the strings in $A$ any number of times. $A^{*}$ also 
includes the null string $\lambda$. For example, if $A = \{xy, z\}$ then, 
$A^{*}$ = $\{\lambda, xy, z, xyz,$ $zxy, xyxy, zz, xyxyxy, xyzxy, ... \}$.
\end{itemize}

In this article, we define regular expressions at a higher granularity for 
the sake of generality. The symbols of the Alphabet $\Sigma$ of our regular expressions are
generic system operations. $\Sigma$ = \{"Access port 1234 of the system," "Overwrite pointer address 
during memory overflow," ..., "Download unwhitelisted malware"\}. All the symbols used in our regular 
expressions can be found in the nodes of the attack graph presented later. These symbols
constitute the regular expression alphabet $\Sigma$.

\subsection{Control-data Flow Graph}
The CDFG of a program is a graphical representation of all possible control paths and data dependencies that the program might encounter during its execution. The basic blocks of the program constitute the nodes of the CDFG.  A basic block is a block of sequential statements that satisfy the following properties:
\begin{itemize}
    \item The control flow enters only at the beginning of the block.
    \item The control flow leaves only at the end of the block.
    \item A block contains a data invariant or a low-level system call. 
\end{itemize}

\begin{figure}[h]
\centering
\includegraphics[width=0.45\linewidth,scale=1]{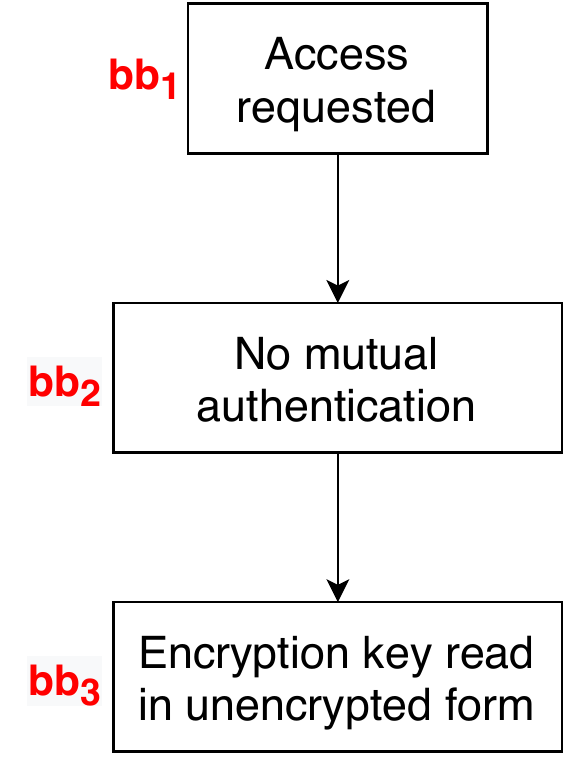}
\caption{Example of a CDFG} 
\label{fig:CDFG_exa}
\end{figure}

Basic blocks are widely used in areas like compiler construction and finite automata. Generally, 
basic blocks denote low-level computer instructions. For the sake of this article, we use basic 
blocks to represent higher-level instructions. We construct the CDFGs at the level of human-executable 
instructions rather than assembly-level instructions, as shown in Fig.~\ref{fig:CDFG_exa}. We do this 
to ensure generalizability of our method to a wide spectrum of applications and systems. 

In Fig.~\ref{fig:CDFG_exa}, $bb_1$ denotes the action of requesting access to a device. Access
can be requested through various ports and protocols like ftp, ssh or http. The port numbers
and protocols to be targeted are system-dependent. We encapsulate these details into a single basic 
block to ensure generalizability of our method to all IoT systems and CPS. Similarly, in 
Fig.~\ref{fig:CDFG_exa}, $bb_3$ denotes the action of accessing the unencrypted key from the device 
key chain of a compromised, rooted or jailbroken device. The exact implementation is 
application-dependent because every application stores its keys at different locations. For example, 
WhatsApp (for both Android and iOS) stores its keys in the file /data/data/com.whatsapp/files/key. 
The assembly-level instructions for accessing the keys are not generalizable since they are 
application-dependent. Hence, we encapsulate all such instructions in a single basic block to 
facilitate generalization of our approach.

\subsection{Support Vector Machine}
We employ ML at the system level. Our training dataset does not have enough training examples to 
train a robust neural network. Thus, we use traditional ML approaches, instead of deep learning, 
for classification. Among traditional ML classification algorithms, 
SVM is one of the most robust classifiers that generalizes quite well.

\par SVM is a class of supervised ML algorithms that analyzes a labeled 
training dataset to perform either classification or regression 
\cite{DBLP:journals/ml/CortesV95}. It is capable of predicting the label of a new example with high accuracy. It is inherently designed to be a linear 
binary classifier. However, kernel transformations can be used
to perform nonlinear classification as well. For a dataset with an 
$n$-dimensional feature space, a trained SVM model learns an $(n-1)$-dimensional hyperplane that serves as the \textit{decision boundary}, also referred to as the \textit{separating hyperplane}.

\par Many contemporary ML algorithms, e.g., $k$-nearest-neighbor 
classification, use a greedy search approach. However, SVM uses a quadratic optimization algorithm to output an optimal decision boundary. The two main limitations of SVM are its natural binding to binary classification and the need to specify (rather than learn) a kernel function.

\section{Methodology}
\label{section::Methodology}
\begin{figure}[h]
\centering
\includegraphics[width=1.05\linewidth,scale=1]{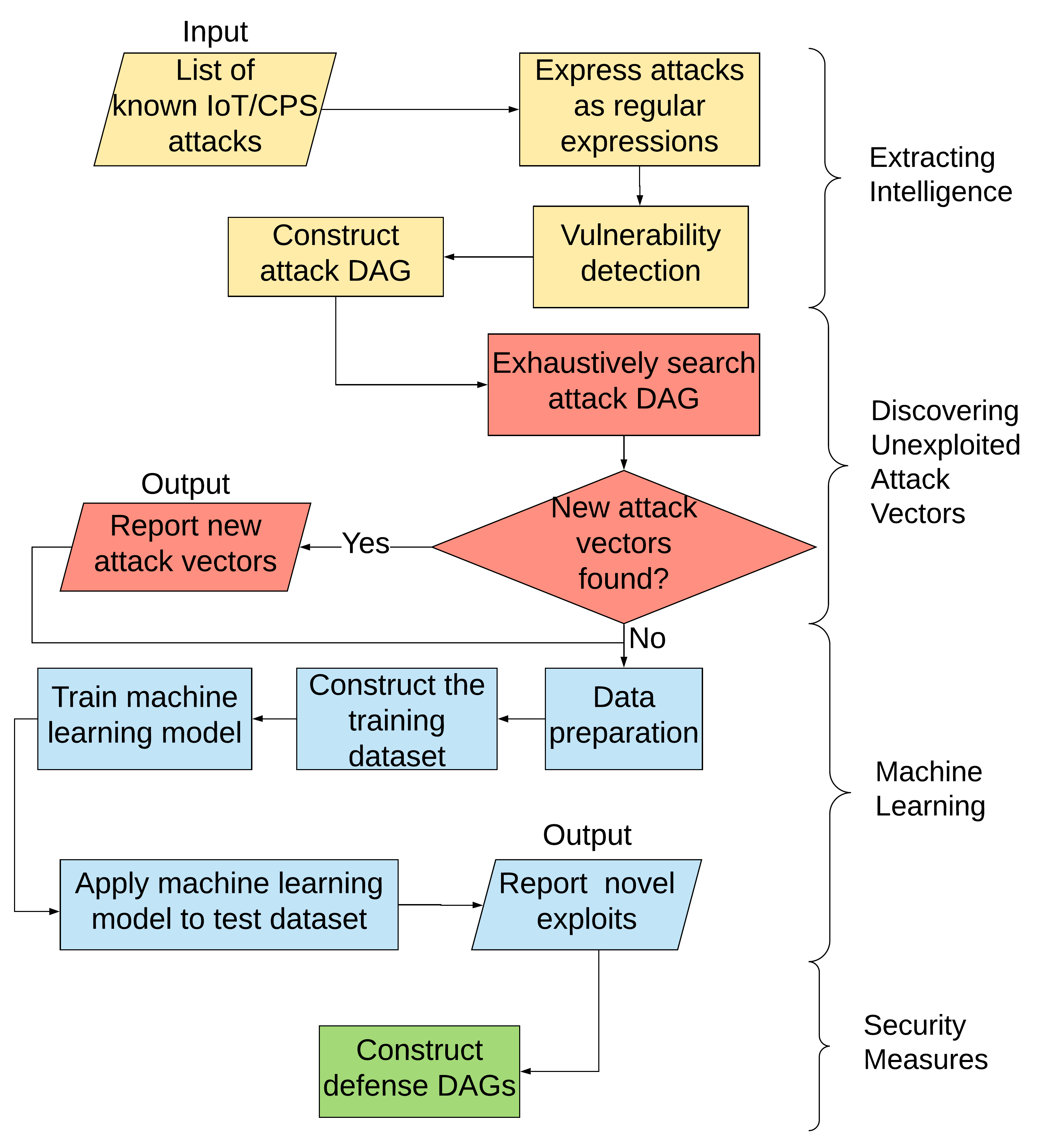}
\caption{Flowchart of the overall methodology} 
\label{fig:flowchart}
\end{figure}

In our methodology, we extract intelligence from an ensemble of known 
CPS/IoT attacks and use this system-level adversarial intelligence to 
predict other possible exploits in a given CPS/IoT framework. The automated derivation of novel exploits and defenses broadly comprises extracting 
intelligence, discovering unexploited attack vectors, applying ML, and taking measures to secure the system. These processes are depicted in the flowchart of 
Fig.~\ref{fig:flowchart}.

\subsection{Extracting Intelligence}
We document existing CPS/IoT attacks and decompose them into their constituent system-level actions and used data invariants. We use regular expressions to 
represent these constituent system-level operations. Then we combine the 
regular expressions of all the attacks to form an ensemble of interconnected system-level operations. This ensemble is represented as a DAG. This DAG is henceforth referred to as the aggregated attack DAG.

\subsubsection{Data Collection}
Next, we discuss how to extract knowledge from known attack patterns. To 
achieve this objective, we create a list of known CPS/IoT 
attacks. Then we classify these attacks into various categories 
based on the type of vulnerability being exploited.  
This list consists of 41 different attacks~\cite{langner2011stuxnet,7924372,mosenia2017comprehensive}.  The most popular attacks among these and their regular expressions are shown in Table~\ref{table:Attacks}.

\subsubsection{Data Transformation}
In this phase, we decompose each attack into its basic system-level operations. 
We express these sequences of operations as regular expressions that are 
then represented as CDFGs.
Each attack is now 
transformed into a CDFG with system-level operations as its basic blocks.
The methodology of decomposing an attack into a CDFG is similar to the method 
used in \cite{DBLP:conf/dimva/AarajRJ08}. 

An example of the data transformation procedure for a buffer 
overflow attack is given next. A buffer overflow attack can be expressed as a sequence of following actions:
\begin{enumerate}
    \item dynamic memory allocation,
    \item overflow of memory, and
    \item frame pointer with overwritten memory.
\end{enumerate}

Let $bb_{i}$ denote the $i^{th}$ basic block of the sequence. Then the
corresponding regular expression is given by:
\begin{eqnarray}
\begin{aligned}
bb_{i}\textit{(Dynamic memory allocation)}^{*}. bb_{j}(\textit{Overflow of memory}).\\
bb_{k}\textit{(Frame pointer with overwritten memory)} \nonumber
\end{aligned}
\end{eqnarray}

\begin{figure}[h!]
\centering
\includegraphics[width=0.4\linewidth,scale=1]{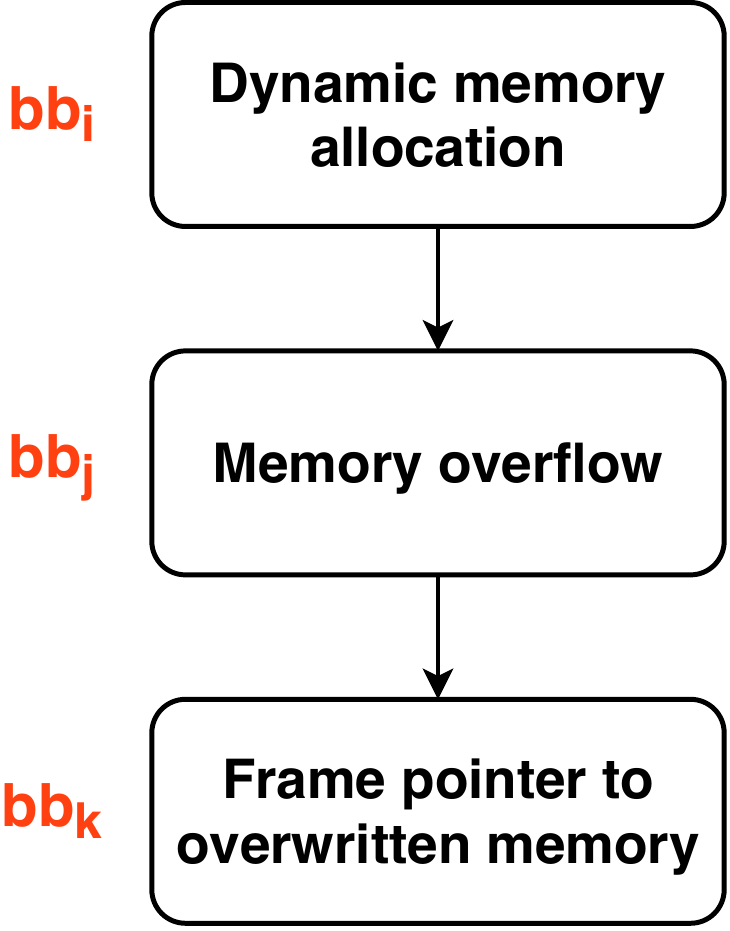}
\caption{CDFG of buffer overflow attacks} 
\label{CDFG_buff_ovrflw}
\end{figure}

Here, $bb_{i}$ denotes the dynamic memory allocation that occurs in the memory  stack before a buffer overflow occurs. The Kleene star operation suggests that $bb_{i}$ might be executed multiple times before $bb_{j}$ is executed. Basic blocks $bb_j$ and $bb_k$ are similarly defined. Every basic block of the regular expression forms a node of the CDFG. The concatenation operation (represented by the dot operator) is represented by a branch from the preceding node of the concatenation operator to its succeeding node. The CDFG of a buffer overflow attack is shown in Fig.~\ref{CDFG_buff_ovrflw}. In Fig.~\ref{CDFG_buff_ovrflw}, we can see that the concatenation between $bb_i$ and $bb_j$ is represented by a directed branch from $bb_i$ to $bb_j$. The Kleene star operator should ideally be represented by a self-loop on the basic block. However, we ignore self-loops in our CDFG representation. This is because we combine these CDFGs into a DAG. The acyclic property of a DAG facilitates our further analysis, but the presence of self-loops violates this property.


\begin{table*}
\begin{center}
\caption{Real-world CPS/IoT attacks and regular expressions}
\begin{tabular}{|M{4cm}|M{3cm}|M{10cm}|}
\hline
\textbf{Attack} & \textbf{Vulnerability category} & \textbf{Regular expression}\\
\hline
Therac-25 Radiation Poisoning & Race condition / TOCTOU vulnerability  & $bb_i$(access system call)*. $bb_j$(open system call)*\\
\hline

Ariane 5 Rocket Explosion & Integer overflow & $bb_i$(data invariant $>$ max integer)* \\
\hline
Worcester Airport Control Tower Communication Hack & Buffer overflow & $bb_i$(dynamic memory allocation)*.$bb_j$(overflow of memory).$bb_k$(frame pointer with overwritten memory)\\
\hline
Bellingham, Washington, Pipeline Rupture & Buffer overflow & $bb_i$(dynamic memory allocation)*.$bb_j$(overflow of memory).$bb_k$(frame pointer with overwritten memory)\\
\hline
Maroochy Shire Wastewater Plant Compromised & Access control/Privilege escalation &
$bb_i$(critical component with one-factor or one-man authentication)*\\
\hline
Davis-Besse Nuclear Power Plant Worm & Malware/Privilege escalation	& $bb_i$(critical
component with one-factor or one-man authentication)* \\
\hline
Worm Cripples CSX Transport System & Malware/Privilege escalation & $bb_i$(critical component with one-factor or one-man authentication)*\\
\hline
Worm Cripples Industrial Plants	& Malware/Privilege escalation & $bb_i$(critical component with one-factor or one-man authentication)*\\
\hline
Browns Ferry Nuclear Plant & Distributed Denial of Service (DDoS) attack & $bb_i$(port traffic per second $>$ threshold)\\
\hline
LA Traffic System Attack & DDoS attack  & $bb_i$(data invariant $>$ threshold)\\
\hline
Aurora Generator Test & Protocol vulnerability & $bb_i$(access requested)*.$bb_j$(no mutual authentication)*\\
\hline
Internet Attack on Epileptics & SQL injection & $bb_i$(user input)*.$bb_j$(user input not compliant with database format)\\
\hline
Turkish Oil Pipeline Rupture & Privilege escalation/ DDoS & $bb_i$(critical component with one-factor or one-man authentication)* + $bb_j$(data invariant $>$ threshold)\\
\hline
Stuxnet Attack on Iranian Nuclear Power Facility & Malware through USB	& $bb_i$(executive file of new executable at kernel level)*.$bb_j$(sending data through port to external C2)\\
\hline
Tests of Insulin Pumps & No authentication + No encryption  & $bb_i$(transaction requested)*.$bb_j$(no time stamp check)*.$bb_k$(no mutual authentication)*.$bb_l$(no hash check)*\\

  & Replay attacks & $bb_i$(data in transit not encrypted)*\\
\hline
Houston, Texas, Water Distribution System Hack & Weak access management
& $bb_i$(access requested)*.$bb_j$(no strong authentication, e.g., no public key infrastructure
based authentication or two-factor authentication)*\\
\hline
Researcher Defeats Key Card Locks & No authentication & $bb_i$(access requested)*.$bb_j$(no mutual authentication)*.$bb_k$(encryption key read from memory in unencrypted format)*\\
\hline
Test of Traffic Vulnerabilities & Weak cryptographic measures &  $bb_i$(no encryption of
data/commands)*+($bb_j$(no digital signature on sensor firmware)*. $bb_k$(illegal access
through unobstructed port)*. ($bb_l$(reconfigure the system specs)* + ($bb_m$(access memory
buffer). $bb_n$(overwrite allocated memory))*))	\\
\hline
German Steel Mill Attack & Malware/Privilege escalation & 	$bb_h$(open downloaded file from spear-phishing email)*.$bb_i$(executive downloaded file from email)*.$bb_j$(critical component with one-factor or one-man authentication)*.$bb_k$(access business network)*.$bb_l$(access ports of entry to production network)*.$bb_m$(manipulate commands to the system)*\\
\hline
Fatal Military Aircraft Crash Linked to Software fault & Software fault & $bb_i$(access system files)*.$bb_j$(rewrite code for updates)*.$bb_k$(delete/modify important system files)*\\
\hline
Test of Smart Rifles & Weak password	& $bb_i$(weak WiFi password)*.($bb_j$(alter state variables)*+$bb_k$(gain root access))\\
\hline
Black Energy Ukrainian Power Grid Attack & Weak authentication& $bb_i$(spear phishing emails to access business network)*.$bb_j$(maneuver into the production network)*.($bb_k$(erased critical files on disk) + $bb_m$(took control over important network nodes)*)\\
\hline
Mirai Botnet Attack & Weak authentication + DDoS & $bb_i$(weak password)*.$bb_j$(port traffic per second $>$ threshold)\\
\hline
Unidentified Water Distribution Facility Hack & Web vulnerabilities & ($bb_i$(phishing emails to access credentials)*+$bb_j$(SQL injection attacks to get credentials)*).$bb_k$(weak storage of credentials on front-end server)	\\
\hline
“WannaCry” Ransomware Attacks & Buffer overflow & $bb_i$(dynamic memory allocation)*.$bb_j$(overflow of memory)*.$bb_k$(frame pointer with overwritten memory in SMBv1 buffer)*\\
 &		Cryptographic key management & $bb_i$(process starts encrypting data)*.$bb_j$(process new to the system and not whitelisted)*\\
\hline
	
\end{tabular}
\label{table:Attacks}		
\end{center}
\end{table*}

\begin{figure*}
\centering
\includegraphics[width=0.799\linewidth,scale=1]{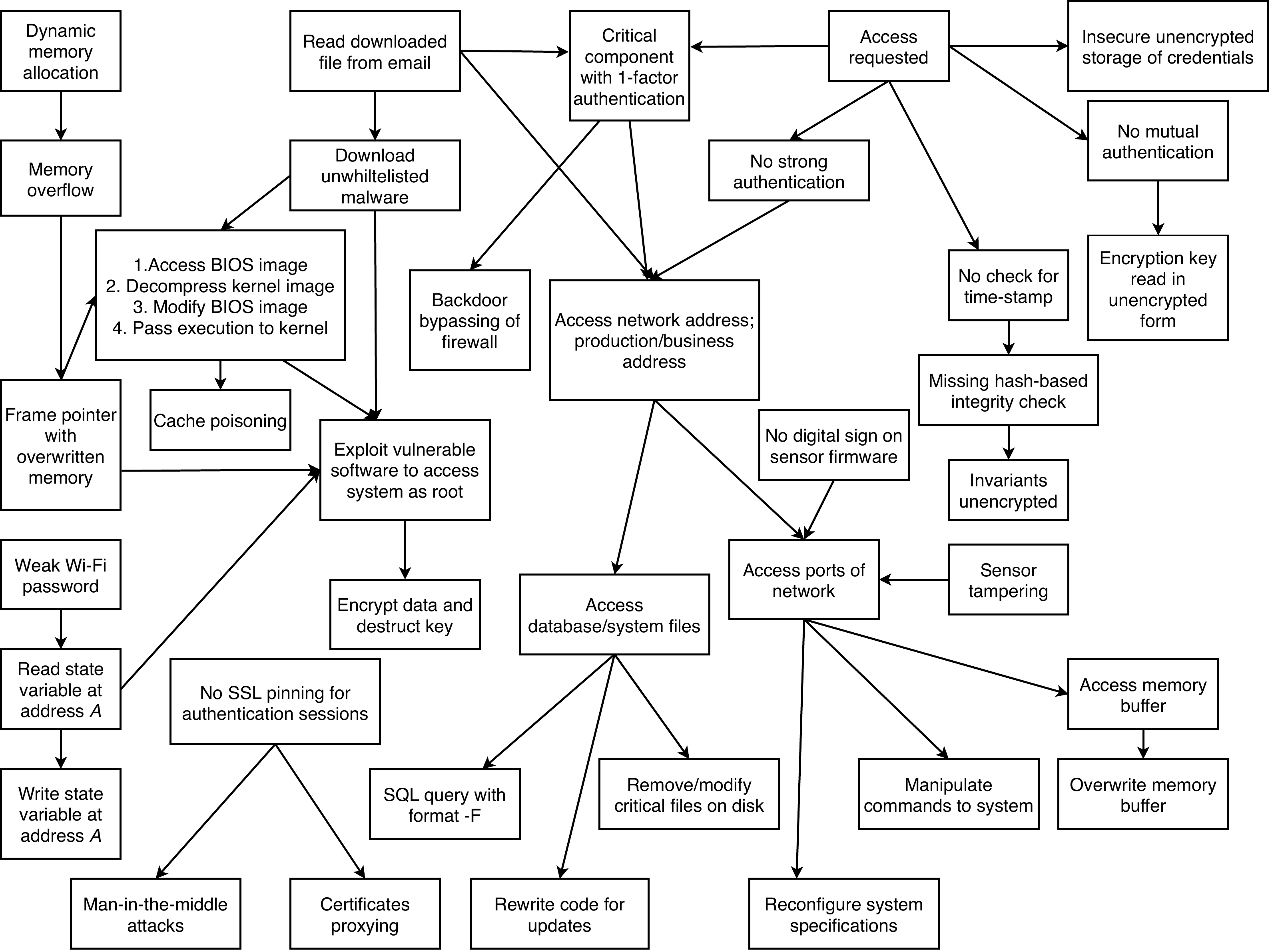}
\caption{The aggregated attack DAG} 
\label{AttackDAG}
\end{figure*}

\subsubsection{Attack DAG}
\label{subsubsection:Attack_DAG}
Every attack in our list is represented by its corresponding CDFG. All the 
CDFGs are combined to form a single DAG. This DAG, shown in Fig.~\ref{AttackDAG}, is our aggregated attack DAG. The attack DAG is a concise representation of the system and network-level 
operations of known categories of CPS/IoT attacks. Every path from a head node to a leaf node in the attack DAG corresponds to a unique attack vector.

We observe that certain basic blocks appear in multiple attacks.
These basic blocks are represented as a single node in the attack DAG with 
in-degree and/or out-degree greater than 1. Our attack DAG has 37 nodes,
represents 41 different attacks, and has a maximum depth of 6.

\subsection{Applying Machine Learning}
Once we have represented the known attacks in the attack DAG, we observe 
that some of its unconnected nodes can be linked together. 
Every new feasible link that is predicted by the ML model is considered to be a novel exploit of vulnerabilities. A link or branch is considered to be feasible if the control/data flow represented by that branch can be implemented in a real-world system. We use ML models to predict if directed branches between various pairs of nodes of the attack DAG are feasible. Manual verification of the feasibility of all possible branches in the attack DAG is too time-consuming.  Let $n$ be the number of nodes in the attack DAG and $c$ be the number of examples in the training dataset. Then the size of the search space of possible branches is
\begin{eqnarray}
    2{n \choose 2} - c &=& n(n-1) - c \nonumber \\
    &=& n^{2} - n - c \nonumber \\
    &=& \Theta(n^{2})
\label{eqn:search_space}
\end{eqnarray}
This quadratic dependence makes it very expensive to perform manual checks to 
exhaustively examine the feasibility of all the possible branches. In our 
experiments, we show that using ML can reduce the search space by 87.2\%.

We train the ML model using the attack DAG of known attack vectors. Once 
trained, it can predict the feasibility of new branches in the attack DAG. 
We derive an SVM model for this purpose. Since the dataset is very small, 
consisting of just 140 datapoints, it prevented us from being able to
adequately train a neural network \cite{ghasemi2018neural}. However, when our 
methodology is applied to a larger scope of cyberattacks, a neural network 
model might be an effective tool~\cite{hassantabar2019scann}.

\subsubsection{Data Preparation}
We assign various attributes (features) to the basic blocks of the attack DAG  depending on the type of impact the attack would have on the system and network. The exhaustive set of attributes that we used is composed of memory, data/database, security vulnerability, port/gateway, sensor, malware, authentication vulnerability, head node, leaf node, and mean depth of the node. Each node has a binary value (0 or 1) associated with every feature except the mean depth. The mean depth of a node denotes the average depth of the node in the attack DAG. For example, nodes \textit{"Certificate proxying"} and \textit{"SQL query with format -F"} have the attributes shown in Table~\ref{table:node_attri}.

\begin {table}[h!]
\begin{center}
\caption {Node attributes}
\begin{tabular}{|c|M{1.3cm}|M{2.2cm}|}
\hline
\textbf{Attribute} & \textbf{Certificate proxying} & \textbf{SQL query with format -F}\\ \hline
Memory & 0 & 0\\
Data/Database & 0 & 1\\
Security vulnerability & 1 & 0\\
Port/Gateway & 0 & 0\\
Sensor & 0 & 0\\
Malware & 0 & 0\\
Authentication vulnerability & 1 & 0\\
Head node & 0 & 0\\
Leaf node & 1 & 1\\
Mean depth & 1 & 3.75\\
\hline
\end{tabular}
\label{table:node_attri}		
\end{center}
\end {table}

\begin{figure}[!h]
\centering
\includegraphics[width=0.8\linewidth,scale=1]{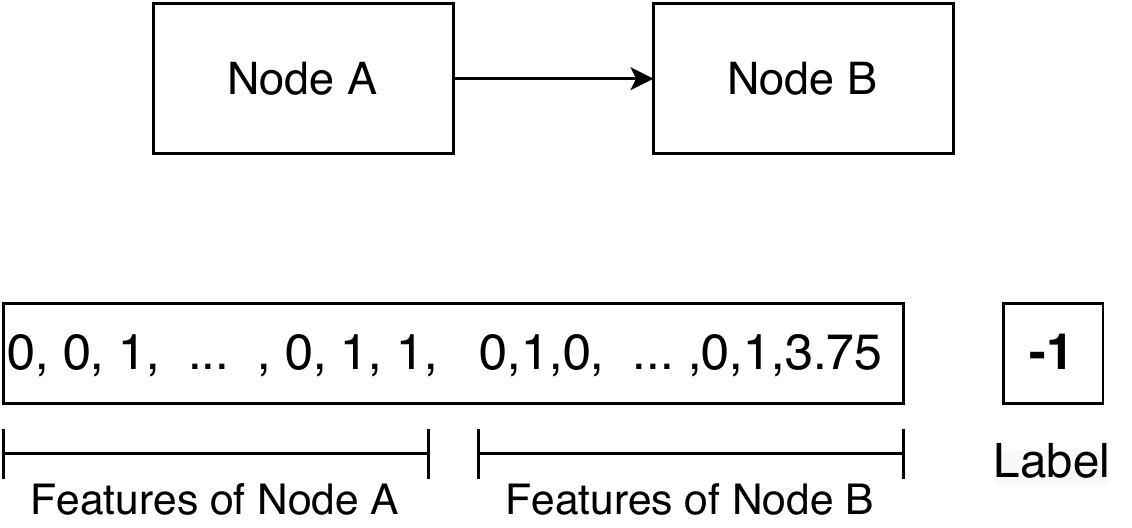}
\caption{Representing a plausible branch as a datapoint of the training or test dataset}
\label{fig:Datapoint}
\end{figure}

We represent a branch in the attack DAG by an ordered pair of nodes, 
i.e., (\textit{origin node, destination node}). The features of the branches 
of the attack DAG are required to train the ML model. The concatenation of the attributes of the origin and 
destination nodes represents the feature vector of a branch. For example, from Table
\ref{table:node_attri}, it can be observed that the feature vector for the node
"\textit{Certificate proxying}" is $[0,0,1,0,0,0,1,0,1,1]$ and that of node \textit{"SQL query
with format -F"} is $[0,1,0,0,0,0,0,0,1,3.75]$. A plausible branch from  node
"\textit{Certificate proxying}" to node \textit{"SQL query with format -F"} will be represented
by the ordered concatenation of the feature vectors of the individual nodes, which is 
$[0,0,1,0,0,0,1,0,1,1,0,1,0,0,0,0,0,0,1,3.75]$. This vector represents the attributes of a plausible 
branch and serves as a datapoint for our ML model. The label of this particular datapoint is
$-$1 because a certificate proxying vulnerability cannot be exploited to launch an SQL injection 
attack. This process of datapoint construction is illustrated in Fig.~\ref{fig:Datapoint}.

\subsubsection{Training Dataset}
Our SVM model learns from the underlying patterns that exist in known CPS/IoT attacks, some
of which are shown in Table~\ref{table:Attacks}. This knowledge is encoded 
in the attack DAG. Thus, the training dataset is composed of all the existing branches (positive examples) and some infeasible branches (negative examples) of the 
attack DAG. The labels of the training dataset are:

\begin{itemize}
    \item 1, if the branch exists in the attack DAG.
    \item $-$1, if a branch from the origin to the destination node is not feasible.
\end{itemize}

A negatively labeled branch denotes an impossible control/data flow. The detailed procedure for generating negative examples for the training set is discussed in Section~\ref{-ve_exa}.

Our training dataset consists of 140 examples, 39 of which have positive 
labels and the remaining have negative labels.

\subsubsection{Negative Training Examples}
\label{-ve_exa}
Creating the set of negative training examples is more complicated than creating the set of
positive training examples. This is because the absence of a branch in the DAG does not imply
that the branch is infeasible. It implies that the branch was not exploited in any real-world
IoT/CPS hack so far. We describe the process of finding negative examples next.

\par First, we classify the nodes of the DAG into broader vulnerability categories. These categories 
are:
\begin{itemize}
    \item Memory vulnerability
    \item Network protocol vulnerability
    \item Weak cryptographic and authentication measures
    \item Malware
    \item Social engineering
\end{itemize}

We observe that some of these vulnerability classes are independent of each other. The pairs of 
independent categories are listed next:
\begin{enumerate}
    \item Memory vulnerability; network protocol vulnerability
    \item Memory vulnerability; social engineering
    \item Network protocol vulnerability; social engineering
    \item Weak cryptographic and authentication measures; malware with social engineering
\end{enumerate}

Most of the nodes of the mutually independent categories will not have feasible branches between 
them. Such infeasible branches are added to the set of negative training examples. However, there may 
exist some exceptions in which branches between nodes of independent categories are feasible.
For example, the nodes "\textit{Overwrite allocated memory}" and "\textit{Access ports of
network}" belong to the mutually independent attack categories, namely \textit{Memory
vulnerability} and \textit{Network protocol vulnerability}, respectively. However, there exist attack vectors in which the buffer overflow vulnerability can be exploited to obtain the device port assignments for the TCP protocol or to listen to network ports on the device. We 
carefully exclude such branches from our set of negative examples.
Similarly, there exist a few infeasible branches between nodes of non-independent categories as 
well as between nodes of the same category. To include such negative branches into our training set, we experimentally 
observe certain statistical properties of the existing negative examples. Then we use these properties 
to filter out more negative examples. The observations are:

\begin{itemize}
    \item Most branches from head nodes to the leaf nodes are infeasible.
    \item Most branches among leaf nodes are infeasible.
    \item The difference between the mean depths of the source and destination nodes of an infeasible 
branch is either smaller than $-0.09$ or greater than $2.0$.
    \item Most infeasible branches have a high Hamming distance (HD) between the feature vectors of 
the source and destination nodes. The mean HD of feasible and infeasible branches were observed to 
be $2.93$ and $4.30$, respectively.
\end{itemize}

Filtering out the probable negative examples with these observations reduces our search space. Then 
we manually select the negative examples from this reduced search space.

\subsubsection{Training}
The ML model has multiple parameters that can be tuned to achieve optimal performance~\cite{CC01a}. The parameters of the SVM model that we experimentally tuned during training are mentioned below.

\begin{enumerate}
    \item \textbf{Regularization parameter (C):} Regularization is used in ML models to prevent overfitting of the model to the training data. Overfitting causes the model to perform well on the training dataset but poorly on the test dataset. This parameter needs to be fine-tuned to obtain optimal performance of the model. The value of $C$ is inversely proportional to the strength of regularization.
    
    \item \textbf{Kernel:} The kernel function transforms the input vector 
$x_i$ to a higher-dimensional vector space $\phi(x_i)$, such that 
separability of inputs with different labels increases. We use the radial basis function (RBF) as our kernel function. The RBF kernel is defined as:
    
    \begin{equation}
        k(x_i,x_j) = exp(-\gamma||(x_i - x_j)||^2)
    \end{equation}
    
    \item $\mathbf{\gamma}$: Parameter $\gamma$ defines how strong the
influence of each training example is on the separating hyperplane.
Higher (lower) values of $\gamma$ denote a smaller (larger) circle of 
influence.
    
    \item \textbf{Shrinking heuristic:} The shrinking heuristic is used to train the model faster. The performance of our model does not change in the absence of this heuristic.
    
    \item \textbf{Tolerance:} The tolerance value determines the error margin that is tolerable during 
training. A higher tolerance value causes early stopping of the optimization process, resulting in a 
higher training error. A higher tolerance value also helps prevent overfitting.
    
\end{enumerate}

\begin{figure}[h!]
\centering
\subfloat[]{
\centering
\label{fig:param3D}
\includegraphics[width=1.1\linewidth]{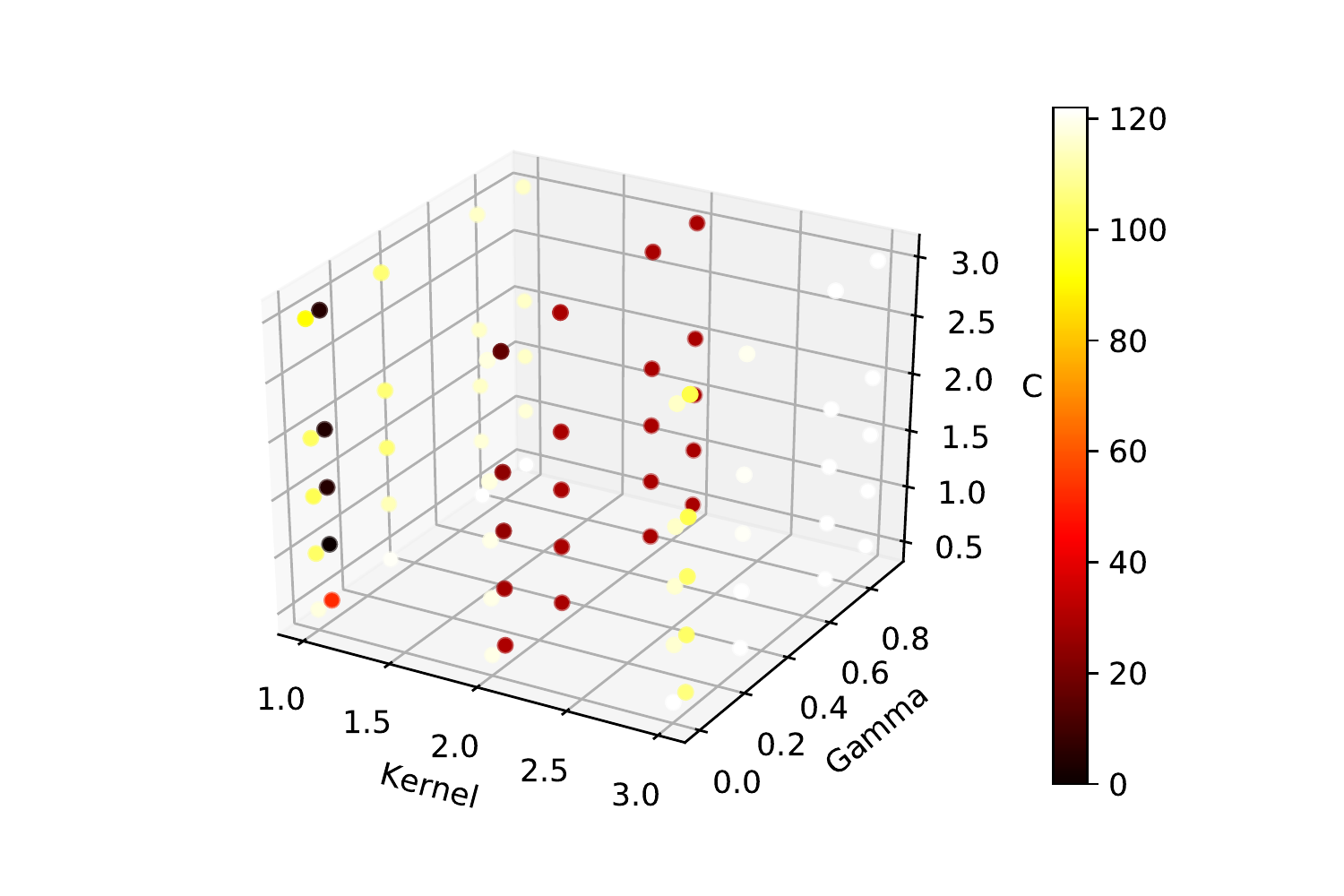}
}

\subfloat[]{
\centering
\label{fig:param_histo}
\includegraphics[width=1.05\linewidth]{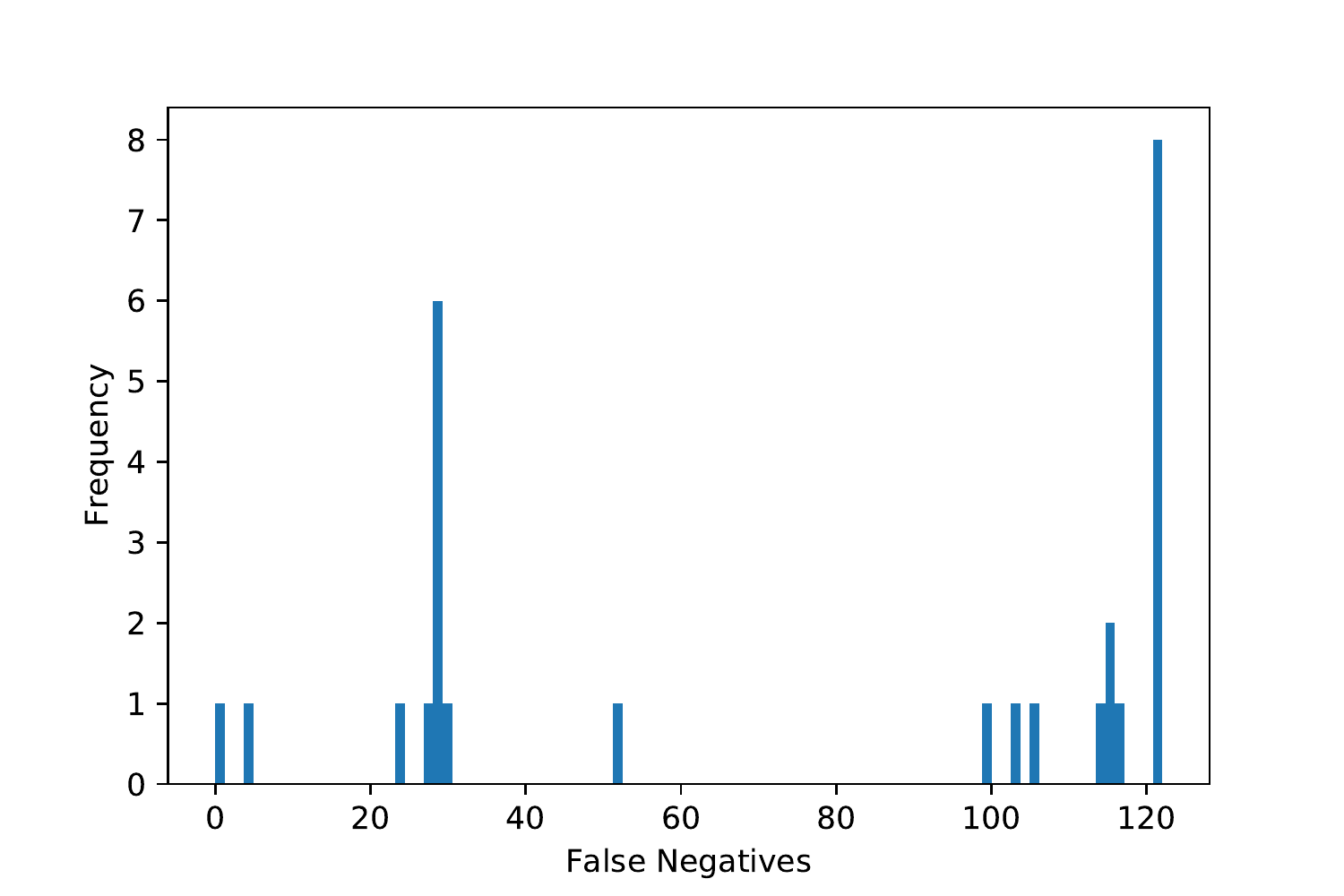}
}

\caption{Parameter search: (a) 3D scatter plot showing the number of FNs for various combinations of 
parameters (Kernels 1, 2, and 3 refer to the RBF, polynomial, and sigmoid kernels,
respectively), and (b) histogram of FN frequencies} 
\label{fig:param_search}
\end{figure}

The parameters of the SVM model are chosen such that we get zero false negatives (FNs). The parameters 
that have the greatest influence on accuracy are the regularization parameter (C), kernel of
SVM, and $\gamma$. We performed a parameter search over various combinations of these parameters and 
plotted the number of FNs against them. The results of our parameter search experiments are shown in 
Fig.~\ref{fig:param_search}. In Fig.~\ref{fig:param3D}, we observe that the RBF kernel yields the lowest number of FNs. We also observe that the number of FNs increases with an increase in the value of $\gamma$. It is very important that we obtain zero FN in order to regard all the negative predictions of the model as infeasible exploits.  In Fig.~\ref{fig:param_histo}, we observe that only one combination 
of parameter values in our parameter search space gives us zero FNs. We choose these parameters for 
our SVM model, as shown in Table~\ref{table:SVM_params}.

\begin {table}[h!]
\setlength{\textfloatsep}{4pt}
\begin{center}
\caption {SVM parameters}
\begin{tabular}{|c|c|}
\hline
\textbf{Parameter} & \textbf{Value}\\ \hline
C & 1.0 \\
Kernel & RBF\\
$\gamma$ & 0.0556\\
Shrinking heuristic & Used\\
Tolerance for stopping & $10^{-3}$\\
\hline
\end{tabular}
\label{table:SVM_params}		
\end{center}
\end {table}

\subsubsection{Verification}
A test example is positive if the sequence of two basic blocks is a permissible control/data flow 
in a given system. Determining the control/data flow in a program is generally a hard task. However, 
in this article, we define the basic blocks at a human-interpretive level. This makes it easier for 
a human expert to determine if the sequence of basic blocks in the test example is feasible or not. 

The SVM model predicts 153 positive labels out of 1192 test datapoints. A positive label indicates 
that the test datapoint is a potential novel exploit. Manual verification of all the 1192 datapoints 
in the test dataset revealed that 1161 predictions by the SVM model are accurate, resulting in a test 
accuracy of 97.4\%.

The parameters of SHARKS were chosen to achieve zero FN. However, our SVM model outputs a few
false positives (FPs). To eliminate these FPs, manual verification is necessary. In the
absence of SHARKS, a human expert would have to verify all 1192 potential vulnerability
exploits manually. With the assistance of SHARKS, it is sufficient to verify only the 153
positive predictions of the SVM model. Thus, SHARKS helps reduce the search space of possible
novel exploits from 1192 to 153, which is an 87.2\% reduction in manual checks.

\subsection{Experimental Results}
In this section, we present the experimental results.
We begin by demonstrating why we chose an SVM model for novel exploit
detection. In addition to SVM, we evaluated the following models: 
k-nearest neighbors (k-NN), naive Bayes, decision tree, and stochastic
gradient descent (SGD) based linear SVM. We compare their accuracies,
precision/recall values, false positive rates (FPR), and F1 scores in 
Table~\ref{table:Results_ML}. It is clear that SVM (with C=1) performs the best.

\begin {table}[h]
\begin{center}
\caption {Performance of ML models}
\begin{tabular}{|M{2cm}|M{1.0cm}|M{1.0cm}|c|c|c|}
\hline
\textbf{Model} & \textbf{Accuracy} & \textbf{Precision} & \textbf{Recall} & \textbf{FPR} & \textbf{F1}\\ \hline
Decision Tree & 86.8\% & 0.40 & 0.89 & 0.14 & 0.55\\ \hline
k-NN (k=2) & 92.8\% & 0.60 & 0.62 & 0.04 & 0.61\\ \hline
k-NN (k=3) & 92.0\% & 0.54 & 0.88 & 0.08 & 0.67\\ \hline
k-NN (k=4) & 94.5\% & 0.70 & 0.70 & 0.03 & 0.70\\ \hline
k-NN (k=5) & 93.0\% & 0.58 & 0.86 & 0.06 & 0.69\\ \hline
Naive Bayes & 90.5\% & 0.46 & 0.26 & 0.03 & 0.34\\ \hline
\textbf{SVM (C=1)} & \textbf{97.4\%} & \textbf{0.80} & \textbf{1.0} &
\textbf{0.03} & \textbf{0.89}\\ \hline
Linear SVM with SGD & 90.6\% & 0.49 & 0.75 & 0.08 & 0.59\\ \hline
SVM (C=2) & 93.8\% & 0.60 & 0.97 & 0.06 & 0.76\\ \hline
SVM (C=3) & 92.8\% & 0.56 & 0.96 & 0.08 & 0.71\\ \hline

\end{tabular}
\label{table:Results_ML}		
\end{center}
\end {table}

We use the SVM model to predict the feasibility of all plausible branches of the attack DAG. The test 
dataset contains all plausible branches except the branches present in the training dataset. The 
branches are converted into test vectors by the method depicted in Fig.~\ref{fig:Datapoint}. The 
attack DAG has 37 nodes and the training set has 140 examples. Putting $n=37$ and $c=140$ in 
Eq.~(\ref{eqn:search_space}), we observe that the test dataset contains 1192 test vectors. The 
SVM model successfully predicts the existence of 122 new feasible branches in the attack DAG.  Each 
new branch corresponds to a unique novel exploit.

Some of the 122 feasible branches of the attack DAG that were predicted by ML are listed in 
Table~\ref{table:zero_day_results}. These attacks have been chosen to represent the most popular 
vulnerability categories.

\begin {table}[h!]
\begin{center}
\caption {Novel exploits discovered}
\begin{tabular}{|M{5cm}|M{2.5cm}|}
\hline
\textbf{Branch discovered} & \textbf{Vulnerability category}\\ \hline
Read downloaded file from email $\longrightarrow$ Overflow of memory &
Buffer overflow \\ \hline
Access network ports $\longrightarrow$ Encrypt data and destroy key & Privilege escalation\\ \hline
Access system files and databases $\longrightarrow$ Reconfigure system specifications& Access control\\ \hline
Download unwhitelisted malware $\longrightarrow$ Bypass firewall using backdoor& Malware\\ \hline
Access network address $\longrightarrow$ Encryption key read from memory in unencrypted form & Cryptographic flaw\\ \hline
Critical component with 1-factor authentication $\longrightarrow$ Access Basic Input/Output System (BIOS) image & BIOS boot level attack\\ \hline
Exploit malware to access system as root $\longrightarrow$ Cache poisoning & Cache poisoning \\ \hline
\end{tabular}
\label{table:zero_day_results}		
\end{center}
\end {table}

The confusion matrix in Table~\ref{table:ConfusionMatrix} shows the number of true negatives
(TNs), FPs, FNs, and true positives (TPs).  The SVM model achieves zero FNs, 
which indicates that a negative prediction is always correct.

\setlength{\textfloatsep}{3pt}
\begin {table}[h!]
\begin{center}
\caption {Confusion matrix}
\begin{tabular}{|c|c|c|c|}
\hline
{N=1192} & \textbf{Actual = No} & \textbf{Actual = Yes} & \\
\hline
\textbf{Predicted = No} & TNs = 1039 & FNs = 0 & 1039\\
\hline
\textbf{Predicted = Yes} & FPs = 31 & TPs = 122 & 153\\
\hline
 & 1070 & 122 & \\
\hline
\end{tabular}
\label{table:ConfusionMatrix}		
\end{center}
\end {table}

\begin{figure}[h!]
\centering
\includegraphics[width=1.1\linewidth,scale=1]{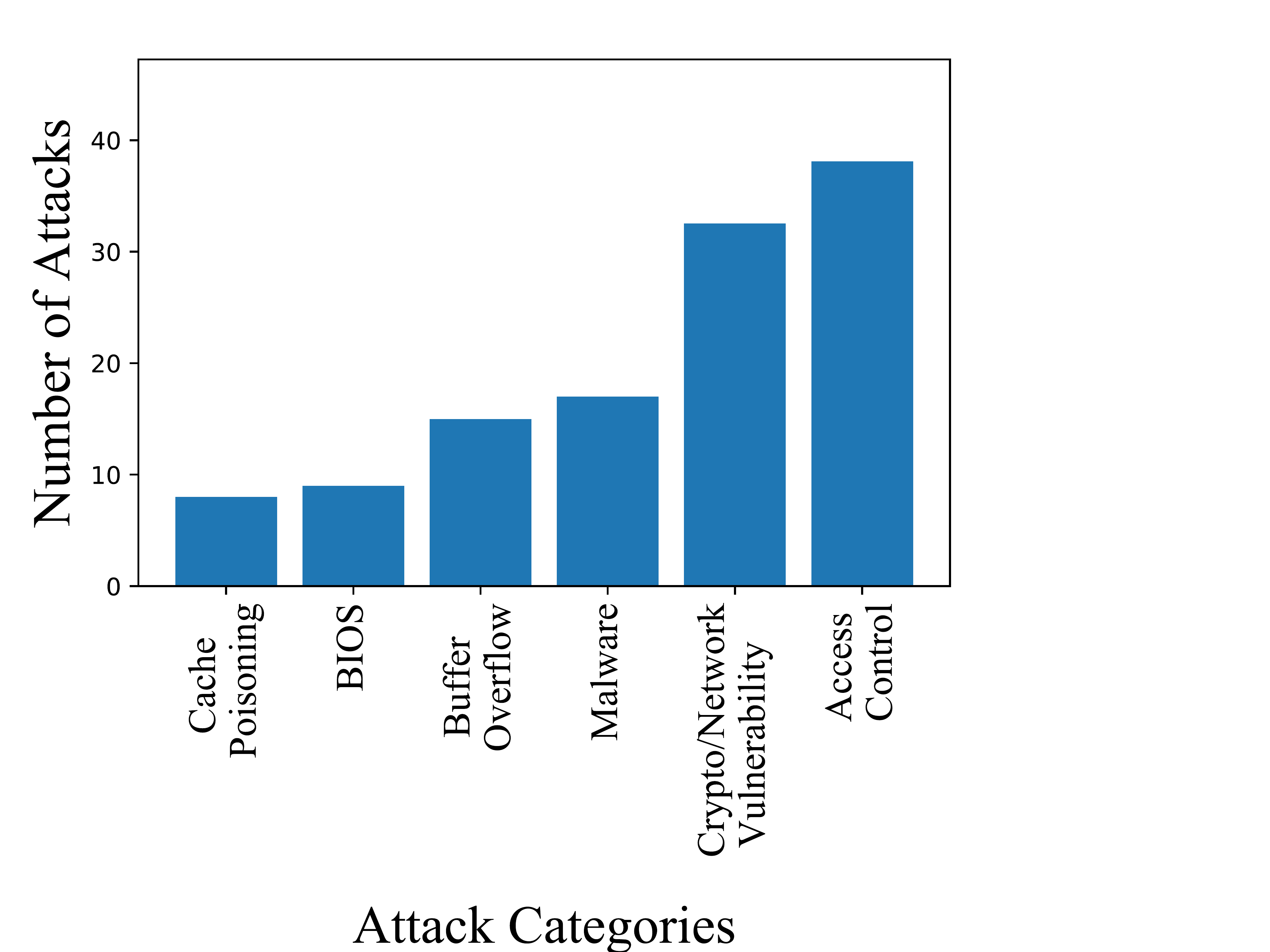}
\caption{Histogram depicting the number of novel exploits discovered in each category} 
\label{fig:histo_attacks}
\end{figure}

In Fig.~\ref{fig:histo_attacks}, we categorize the novel exploits into 
six categories. We can see that access control vulnerabilities (including 
privilege escalation), weak cryptographic primitives, and network security 
flaws are most common vulnerabilities with high likelihood of exploit. We also observe that vulnerabilities with lower exploit likelihood are BIOS vulnerabilities and cache poisoning 
attacks. This is expected because a successful BIOS attack or a cache 
poisoning attack involves one or more of the following: boot-stage execution, shared resources with adversary, side-channel access, kernel code execution, and close proximity to the CPS/IoT devices at very specific time instances. These operations involve higher complexity in building the exploit chains across various system elements.

Training accuracy refers to the accuracy of the SVM model when evaluated on the training
dataset. Only four of the 140 training datapoints were incorrectly classified by the SVM model,
yielding an accuracy of 97.2\%. The test accuracy is manually determined by evaluating the feasibility of all the 1192 possible branches in the attack DAG. We observed that 31 of the 153 positive predictions were incorrect. On the other hand, all the negative predictions were accurate. Thus, 1161 of the 1192 datapoints of the test dataset were classified correctly by the SVM model, yielding a test accuracy of 97.4\%.

\subsubsection{Constraint Satisfaction Problem (CSP) Formulation}
In this section, we use a CSP formulation to analyze the feasibility of DAG branches. 
Constraint-based reasoning requires the creation of a set of constraints over the features of the 
branches, such that any branch satisfying all the constraints is deemed to be a feasible branch. 
CSP is widely used in program analysis techniques. Unlike traditional CSP-addressable problems, 
there are no deterministic rules governing the feasibility of a branch in our attack DAG. Thus, 
generating mathematical constraints that define the feasibility of a branch is not easy.

We use heuristics derived from the statistics of the training set to generate constraints. The 
constraints used to detect infeasible branches are inspired by the thresholds derived to construct 
the negative examples in Section~\ref{-ve_exa}.  The variables and symbols are defined next:
\begin{itemize}
    \item Hamming distance - $hd$ (integer value)
    \item Height difference of nodes - $ht_{diff}$ (float value) 
    \item Head-leaf connection - $hl$ (Boolean value)
    \item Leaf-leaf connection - $ll$ (Boolean value)
\end{itemize}

Experimentally, the following results were observed on the training set.

\begin{enumerate}
    \item Mean Hamming distance between feature vectors of nodes of feasible branches = 2.93.
    \item Mean Hamming distance between feature vectors of nodes of infeasible branches = 4.30.
    \item Height difference between nodes of feasible branches: Minimum = -0.08; Mean = 0.997;
Maximum = 2.
    \item Height difference between nodes of infeasible branches: Minimum = -3.33; Mean =
0.071; Maximum = 3.33.
    \item (Number of infeasible head-leaf or leaf-leaf branches/ Number of feasible head-leaf
or leaf-leaf branches) = 4.
\end{enumerate}

These observations are used to generate the following heuristic constraints. A branch is deemed to be infeasible if:

\begin{enumerate}
    \item $ht_{diff} \leq -0.09$ or $ht_{diff} >2$
    \item $hd > 5$
    \item if $(4 \leq hd \leq 5)$ and $((hl == \text{True}) or (ll == \text{True}))$
\end{enumerate}

When the above constraints are applied to the test dataset, they yield 60 FNs. Comparing this to the 
results in Fig.~\ref{fig:param_histo}, we observe that CSP analysis performs better than many SVM 
models. However, the best SVM model yields zero FNs, thus performing significantly better than CSP. 
Since our primary objective is to minimize the number of FNs, we prefer SVM over CSP.

\subsection{Discovering Unexploited Attack Vectors}
\label{section:exhaustive_search}
\begin{figure}[h!]
\centering
\subfloat[]{
\centering
\label{fig:CDFG1}
\includegraphics[width=0.65\linewidth]{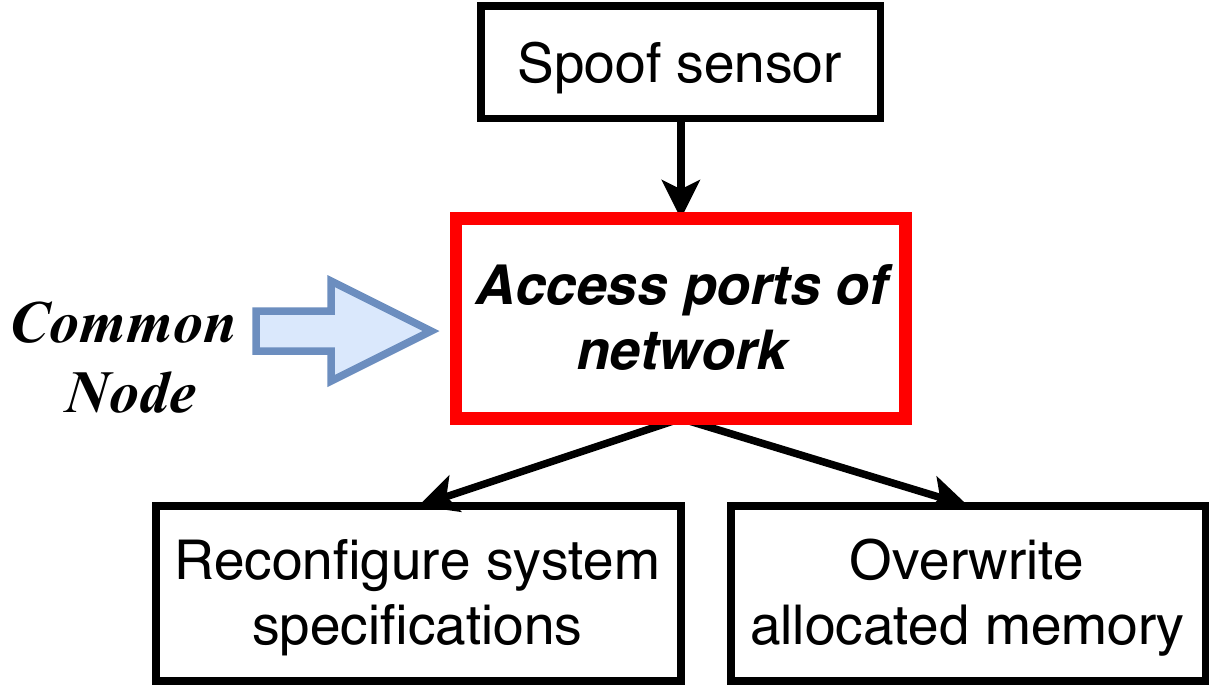}
}

\subfloat[]{
\centering
\label{fig:CDFG2}
\includegraphics[width=0.65\linewidth]{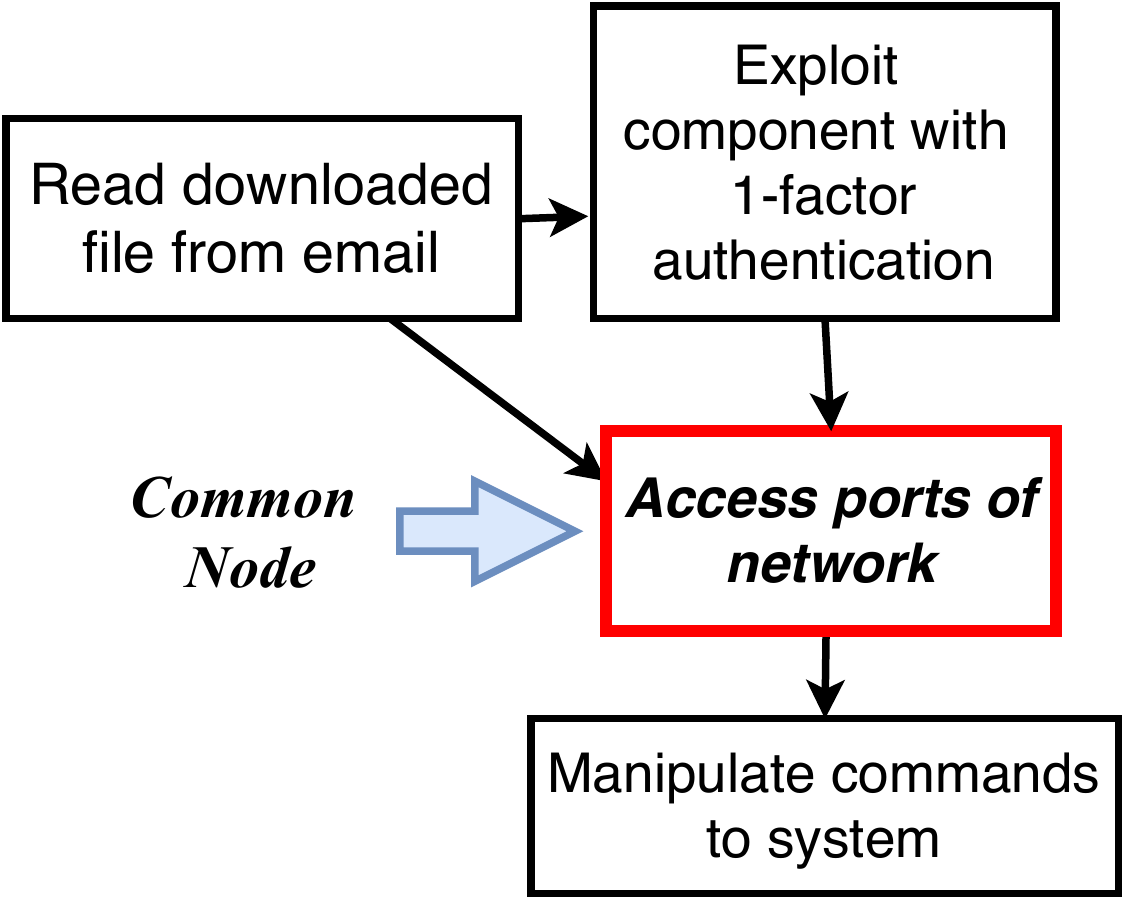}
}

\subfloat[]{
\centering
\label{fig:CDFG3}
\includegraphics[width=0.9\linewidth]{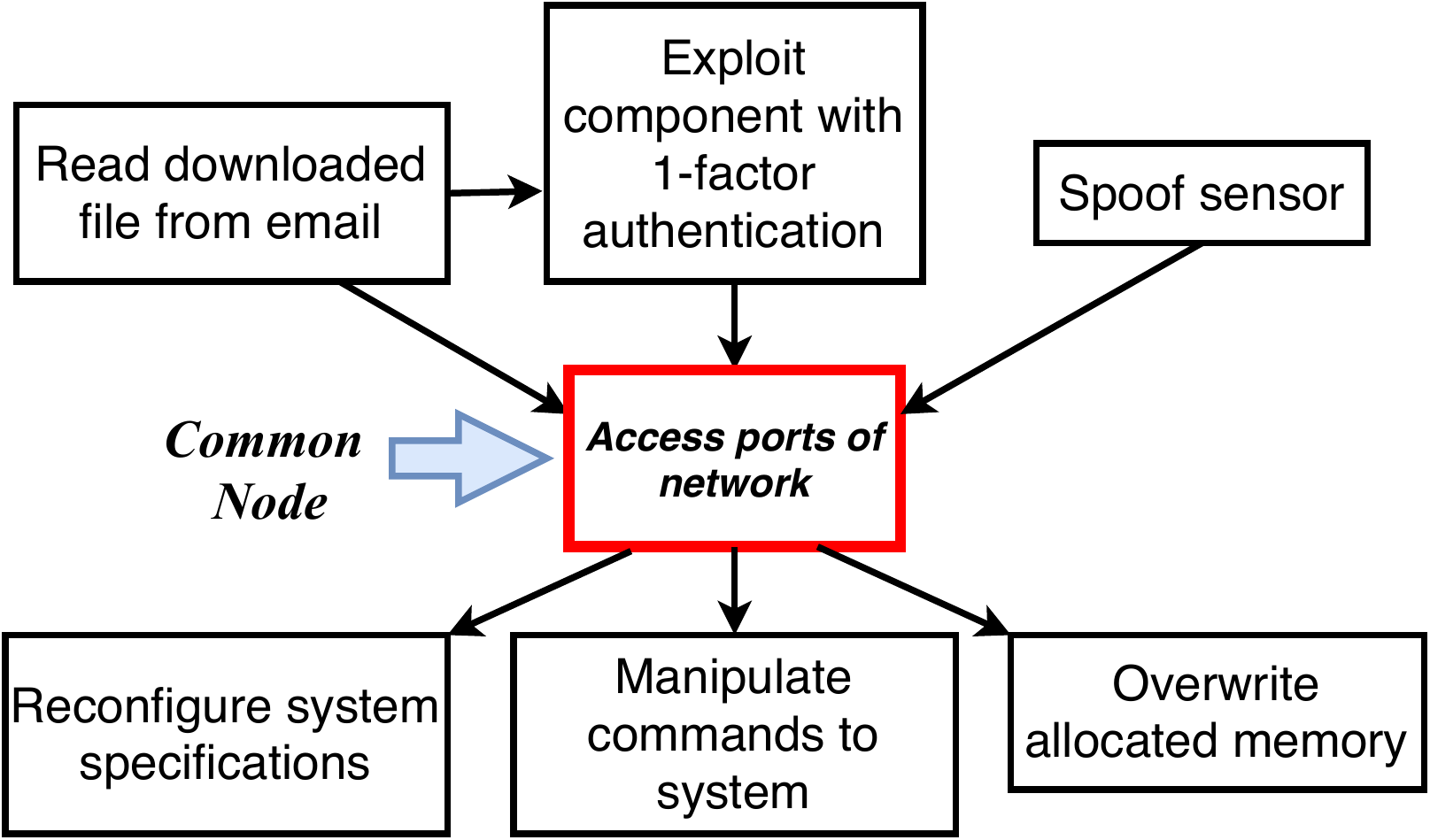}
}

\subfloat[]{
\centering
\label{fig:CDFG4}
\includegraphics[width=\linewidth]{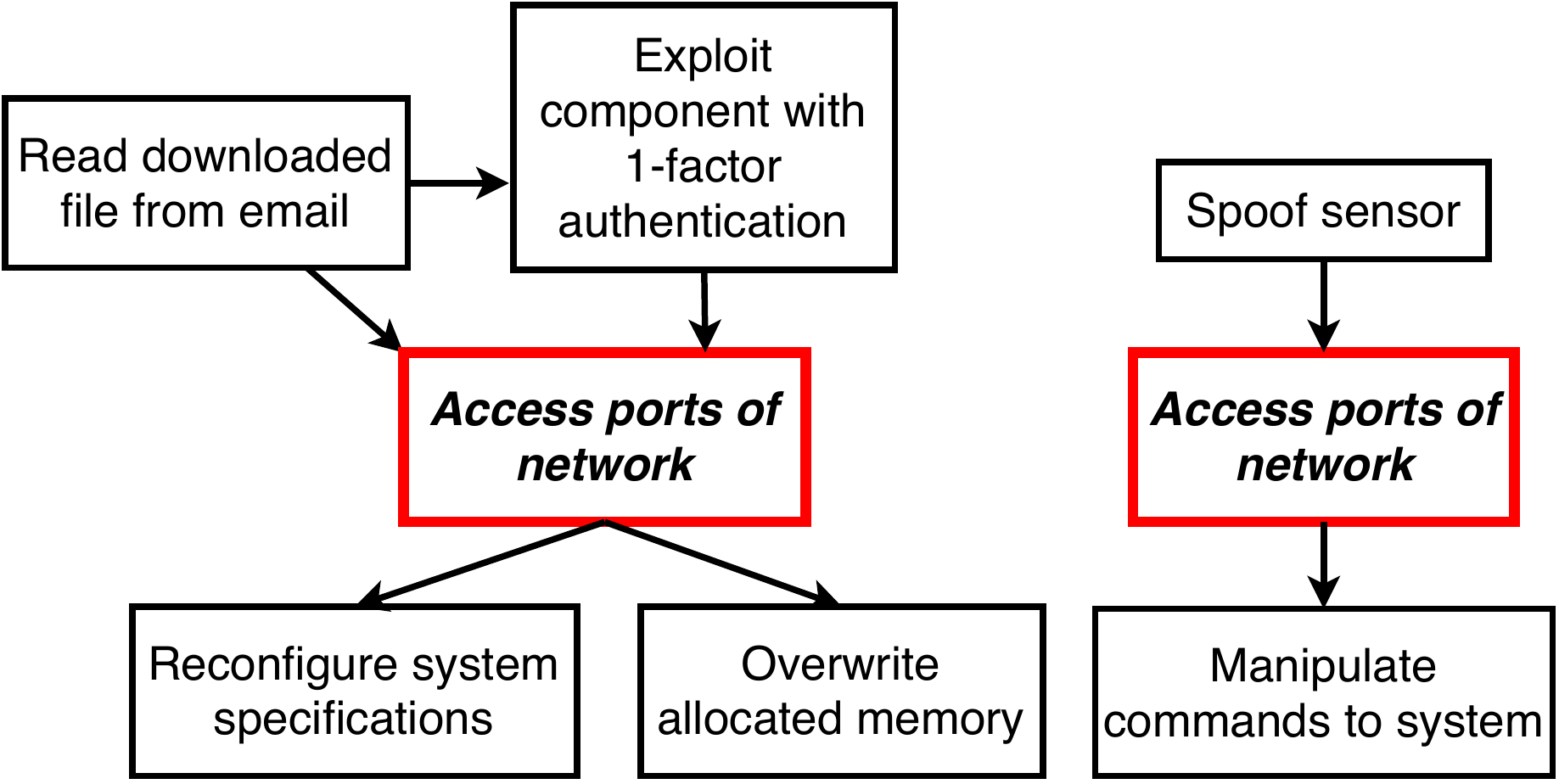}
}
\caption{Generating new exploits with linear search: (a) CDFG for 
$Attack_{1}$, (b) CDFG for $Attack_{2}$, (c) combined CDFG of both attacks, 
and (d) new attacks that emerge from a combination of the two CDFGs} 
\label{fig:Exhaustive_search}
\end{figure}

An unexploited attack vector is one that exists in the DAG but has not yet been exploited in any 
documented real-world attack on IoT/CPS. This is unusual because the DAG was constructed from 
real-world attacks on IoT/CPS. The unexploited attack vectors embedded in the DAG can be discovered 
through linear search on the DAG. Every path from a head node to a leaf node corresponds to a unique 
attack vector. The attack DAG has 51 such paths. However, only 41 known attack vectors were 
considered while constructing the attack DAG. Thus, 10 unexploited attack vectors are obtained 
through a linear search of all the attack paths, a subset of which is shown in 
Fig.~\ref{fig:Exhaustive_search}. These 10 unexploited attack vectors represent 10 additional ways 
to compromise an IoT/CPS.

New attack vectors emerge due to the convergence of multiple attack paths at common basic block(s). Such an occurrence is illustrated in 
Fig.~\ref{fig:Exhaustive_search}. Fig.~\ref{fig:CDFG1} and Fig.~\ref{fig:CDFG2} 
represent two subgraphs of the attack DAG in Fig.~\ref{AttackDAG}. 
Fig.~\ref{fig:CDFG3} shows the graph obtained by combining 
Fig.~\ref{fig:CDFG1} and Fig.~\ref{fig:CDFG2} at the common node titled 
``\textit{Access ports of network}." Fig.~\ref{fig:CDFG4} depicts the new 
paths obtained from the combination of the two graphs. The five new paths thus 
discovered correspond to five attack vectors that have not yet been 
exploited in real-world CPS/IoT attacks.

\section{IoT Case Study: Connected Car}
\label{section:Case_study}
\begin{figure}[h!]
\centering
\includegraphics[width=1.06\linewidth,scale=1]{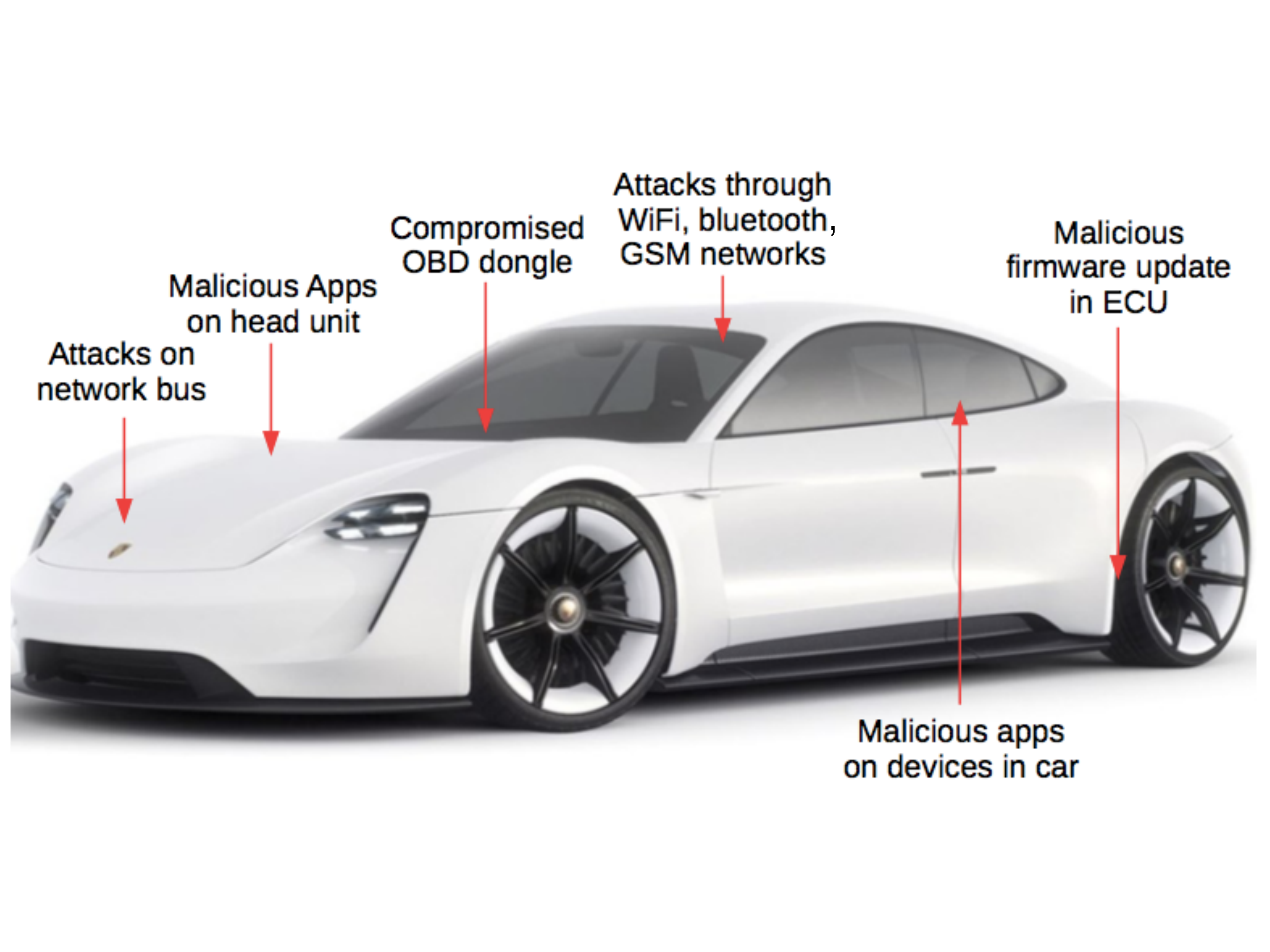}
\caption{Examples of hacker entry points into the connected car} 
\label{fig:smart_car}
\end{figure}

The connected car is a complicated IoT system comprising various 
sensors, electronic control units (ECUs), system buses, and embedded software 
packages. It possesses a vast range of capabilities that includes Internet 
access, communication with multiple devices, and collection of real-time data 
from the surroundings. While these functionalities enhance user convenience, 
they also expand the attack surface of the system. The attack surface of a connected car includes 
the networks of vehicle-to-vehicle communication, in-vehicular communication, and exposed software 
and sensors, to name a few~\cite{lima2016towards}. We analyze the security of in-vehicular networks 
with the SHARKS framework. The most common entry points 
into the in-vehicular network are the ECUs, on-board diagnostics (OBD) port, WiFi, and GSM and 
bluetooth networks of the vehicle. Some of these are shown in 
Fig.~\ref{fig:smart_car}. 

The connected car has numerous ECUs that are responsible for different 
functionalities like anti-lock braking, lane departure warning, and engine 
management. All communications between ECUs occur over the network bus that connects all the ECUs 
to one another. There exist multiple networks for in-vehicle communications. Some of them are local 
interconnect network~\cite{ruff2003evolution}, FlexRay network~\cite{makowitz2006flexray}, 
and media-oriented systems transport network~\cite{fijalkowski2011media}. One of the most popular 
in-vehicle networks is the Controller Area Network 
(CAN)~\cite{tuohy2014intra}. CAN ensures real-time handling of all 
in-vehicle communications, including safety-critical data. This makes the 
security of the CAN bus critical to the safety and security of the smart 
vehicle. However, the CAN bus has been shown to be intrinsically 
not secure~\cite{avatefipour2017linking,choi2018identifying}. Cryptographic 
techniques like encryption and message authentication cannot be applied to
the data traversing the CAN bus. These operations increase the latency of 
processing the packets that leads to an increased ECU response time.  This
overhead is not permissible in the case of safety-critical, time-sensitive, 
and real-time applications. Cryptographic measures also prevent car mechanics 
from analyzing CAN traffic during troubleshooting. This is a major 
inconvenience for them because they generally use the CAN bus as a diagnostic 
tool during repair.


\subsection{CAN Bus Vulnerabilities}
\label{section:CAN_vuln}
Although CAN is the de-facto in-vehicle network in connected vehicles, it 
is not secure by design. The CAN protocol uses a broadcast mechanism for 
communication. Due to the absence of sender and receiver addresses in the data 
frames, every ECU can freely publish and receive messages from the bus. While 
this enables easier addition of new ECUs to the network, it poses a grave 
security threat to the system. We next discuss the popular vulnerabilities on the CAN 
bus that were detected by our approach.

\begin{enumerate}
    \item \textbf{Frame sniffing:} The CAN protocol uses a broadcasting 
mechanism for ECU communications. This allows a malicious node on the CAN bus 
to receive all the data frames through sniffing. The absence of encryption 
makes it easier to analyze the collected frames. The range of valid messages on 
the CAN bus is small enough to be exhaustively analyzed. Fuzzing 
techniques can be used to decode the functionalities of various ECUs from the 
log of sniffed frames~\cite{koscher2010experimental}. This is a breach of 
confidentiality of the system. Frame sniffing is often the precursor of more 
complex attacks.
    
    \item \textbf{Frame spoofing:} Frame spoofing involves sniffing and reverse 
engineering of the data frames of the CAN bus. Using the details of the data 
frames, the adversary can broadcast malicious frames on the bus by spoofing a 
particular node. Absence of authentication schemes compromises the integrity 
of messages on the CAN bus. Spoofing attacks may result in incorrect 
speedometer readings, arbitrary acceleration of the vehicle, erroneous fuel 
level readings, and conveying malicious messages to the 
driver~\cite{liu2017vehicle}. This poses grave safety concerns as the 
adversary can gain access to safety-critical ECUs like the braking system and 
engine management system.
    
    \item \textbf{Denial of Service (DoS):} The CAN protocol implements 
a priority-based broadcasting communication scheme. For example, messages from 
the anti-lock braking system, which are critical to the safety of the 
passengers, are given higher priority for transmission on the bus than 
messages from climate control sensors. The priority of a frame is determined 
by a parameter id (PID). Lower values of PID signify higher priority messages. 
To launch a DoS attack, the adversary needs to decode the smallest acceptable 
value of PID from the history of CAN messages (obtained by frame sniffing). 
Then he can continually broadcast messages with the highest priority on the 
bus, thus preventing any other message from being transmitted on 
it~\cite{koscher2010experimental}. This compromises the availability of the 
CAN bus to legitimate messages, thus denying service to these messages. 
    
    \item \textbf{Replay attack:} Replay attacks involve sniffing the frames on 
the CAN bus prior to launching the attack. Sniffing and analyzing the frame 
packets using fuzzing techniques reveal knowledge about the 
frame functionalities. Since the CAN protocol is bereft of authentication 
schemes and time-stamp verification, the recorded frame packets can be sent 
on the CAN bus at inconvenient time instances to launch various attacks. For 
example, the frame packet to unlock the car door can be replayed by a thief 
when the owner is not around. Replay attacks on cars have been demonstrated 
both in simulations~\cite{hoppe2007sniffing} and real 
cars~\cite{koscher2010experimental}.
    
\end{enumerate}

The other vulnerabilities that we consider in our experiments are ECU buffer 
overflows~\cite{checkoway2011comprehensive} and malware injection through ECU 
firmware updates~\cite{nilsson2009conducting}. These attack vectors involve sending malicious 
packets to the ECUs over the CAN bus but do not involve exploiting any vulnerability of 
the CAN bus itself.

\subsection{Application of SHARKS}
In this section, we describe how we use our SHARKS approach to detect the 
aforementioned vulnerabilities in the given IoT system, namely the CAN bus. An adversary can gain access to the CAN bus through multiple entry points like the OBD port, WiFi, bluetooth, radio or the GPS system of the car~\cite{payne2019car}. In our simulations, we use the OBD-II port to gain access to the CAN bus. We simulate the CAN bus with OpenGarages ICSim simulator~\cite{payne2019car} on our workstation with LibSDL and 
Socket-CAN CAN-utils libraries. The simulation results can then be executed on a connected car, with the help of ScanTools software, by connecting the workstation (laptop) to the car through the OBD-II port.

\begin{figure}[h]
\centering
\includegraphics[width=0.82\linewidth,scale=1]{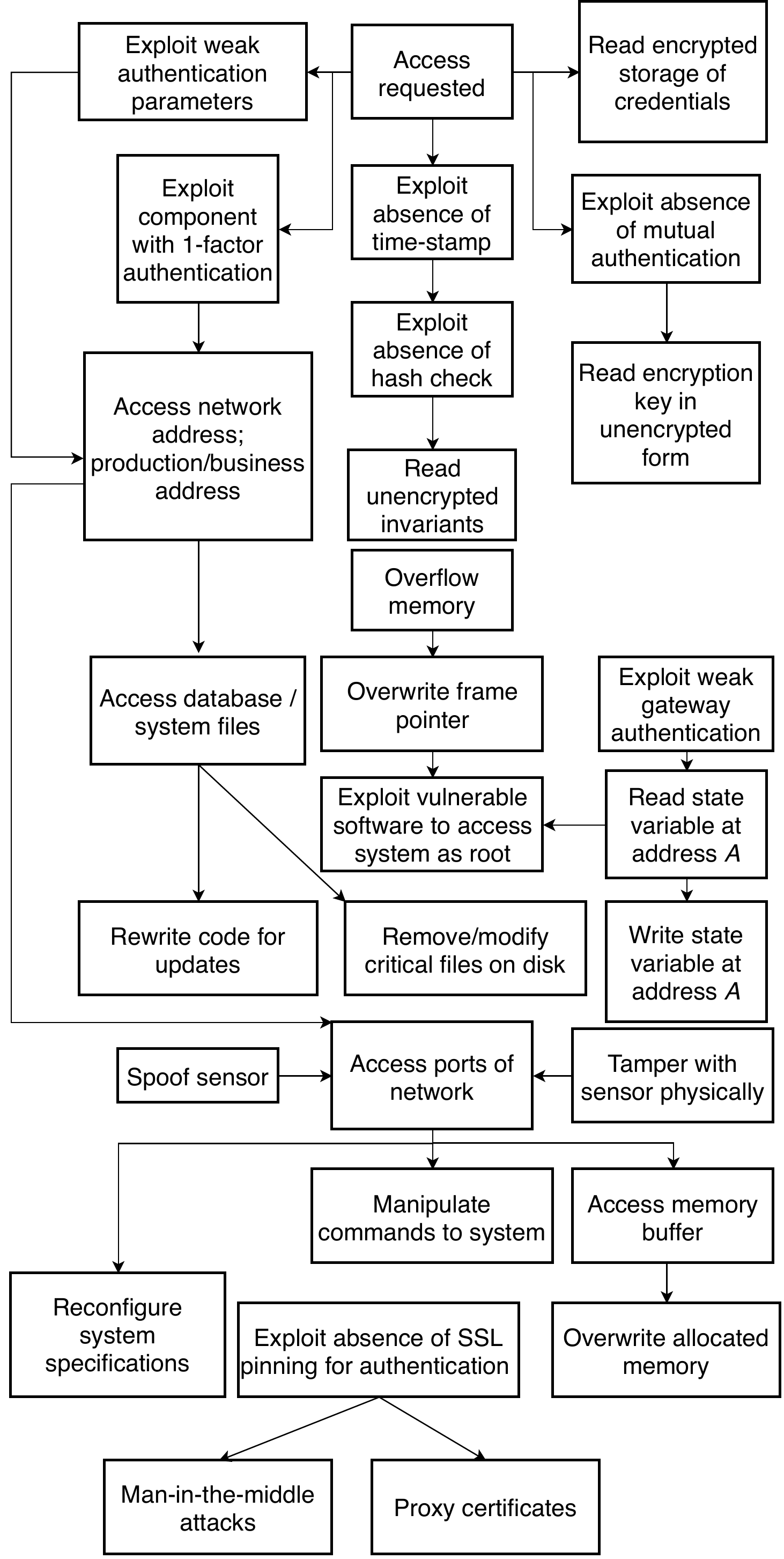}
\caption{The attack DAG for the CAN bus of the connected vehicle}
\label{fig:CAN_DAG}
\end{figure}

To apply SHARKS to a specific CPS/IoT system, we have to design the attack 
DAG for it. The attack DAG shown in Fig.~\ref{AttackDAG} is designed for a 
generic CPS/IoT system. The CAN bus has fewer functionalities than those 
considered during the design of the attack DAG in Fig.~\ref{AttackDAG}. This 
makes some of the nodes in the DAG in Fig.~\ref{AttackDAG} redundant with 
respect to the CAN bus IoT system. We remove those nodes and obtain a subgraph 
of Fig.~\ref{AttackDAG} that is relevant to the CAN bus. This subgraph, shown 
in Fig.~\ref{fig:CAN_DAG}, is referred to as the CAN attack DAG. It has 29 
nodes, 27 branches, and represents 24 high-level attack vectors relevant to the CAN bus.

\subsection{Results}
We ran a pre-trained SVM model on the CAN attack DAG shown in 
Fig.~\ref{fig:CAN_DAG}. The SVM model was trained on the attack DAG in 
Fig.~\ref{AttackDAG}, and not on the CAN attack DAG. While testing the 
model's performance on the CAN attack DAG, we observed that it is able to 
discover 67 CAN vulnerability exploits that were initially absent in the CAN 
attack DAG. This indicates that our approach is generic enough to be deployed on any CPS/IoT system 
for vulnerability exploit detection. We classify the detected CAN bus vulnerabilities into the 
vulnerability categories mentioned in Section~\ref{section:CAN_vuln}. Some of the attack branches 
predicted by the model and their corresponding categories are shown in Table~\ref{table:CAN_results}.

\begin {table}[h!]
\begin{center}
\caption {Novel exploits discovered}
\begin{tabular}{|M{5cm}|M{2.5cm}|}
\hline
\textbf{Branch discovered} & \textbf{Vulnerability category}\\ \hline
Invariants unencrypted $\longrightarrow$ Read state variable at address \textit{A} &
Frame sniffing \\ \hline
No mutual authentication $\longrightarrow$ Man-in-the-middle attacks & Frame spoofing\\ \hline
No check for time-stamp $\longrightarrow$  Manipulate commands to system & Replay attack\\ \hline
Access memory buffer $\longrightarrow$ Write state variable at address \textit{A} & ECU buffer overflow\\ \hline
Rewrite code for updates $\longrightarrow$ Download unwhitelisted software & Malware injection through ECU updates\\ \hline
\end{tabular}
\label{table:CAN_results}		
\end{center}
\end {table}

The CAN attack DAG has 29 nodes and 27 branches. Putting $n=29$ and
$c = 27$ in Eq.~(\ref{eqn:search_space}), we observe that there are 785
datapoints in the test set. The SVM model predicts
88 of these to be feasible novel exploits and eliminates the rest.
Manually examining the feasibility of the 88 positive predictions, we
find that 67 of them are TPs. All the branches that were predicted
to be negative are infeasible control/data flows. Hence, the SVM model reduced our search space from 
785 to 88, which represents an 88.8\% reduction in human effort.  The confusion matrix of the 
predictions made by the model is shown in Table~\ref{table:CAN_ConfusionMatrix}.

\begin {table}[h]
\begin{center}
\caption {Confusion matrix of SHARKS on CAN vulnerabilities}
\begin{tabular}{|c|c|c|c|}
\hline
{N=785} & \textbf{Actual = No} & \textbf{Actual = Yes} & \\
\hline
\textbf{Predicted = No} & TN = 697 & FN = 0 & 697\\
\hline
\textbf{Predicted = Yes} & FP = 21 & TP = 67 & 88\\
\hline
 & 718 & 67 & \\
\hline
\end{tabular}
\label{table:CAN_ConfusionMatrix}		
\end{center}
\end {table}

\section{Security Measures}
\label{section:Security}
In this section, the primary endeavor is to show how to defend CPS/IoT against the known attacks 
and novel exploits predicted by SHARKS at an optimal cost.  Defense-in-depth and multi-level 
security (MLS)~\cite{DoD_5200.28_1,DoD_5200.28_2} are the most appropriate schemes to adopt in 
such a scenario. Defense-in-depth refers to employing multiple defense strategies against a single 
weakness and is one of the seven properties of highly secure devices \cite{hunt2017seven}. 
MLS categorizes data/resources into one of the following security levels: \textit{Top Secret}, 
\textit{Secret}, \textit{Restricted}, and \textit{Unclassified}. 
The first three levels have classified resources and require different levels of protection. 
Security measures become stricter as we move from Restricted to Top Secret. Many different policies 
can be employed to implement MLS in an organization. Some of the most popular policies are based on 
the Bell-La Padula (BLP) model~\cite{rushby1986bell} and the Biba model~\cite{biba1977integrity}. The 
BLP model prioritizes data confidentiality whereas the Biba model gives more importance to integrity.

The aggregated attack DAG is composed of multiple categories of attacks that are weaved together. 
Defense mechanisms can be systematically developed for each of these vulnerability categories in 
the form of defense DAGs. Defense DAGs mirror the corresponding attack subgraphs and make execution 
of the key basic blocks of the attack sequence infeasible. This ensures that no path from a head node 
to a leaf node in the attack DAG can be traversed in the presence of the suggested defense measures.

Many attacks have multiple defense strategies that can protect against them. The cost of our overall 
defense strategy increases with the complexity and number of defense measures that we enforce. 
Defense-in-depth helps us optimize this cost. The less sensitive resources (those belonging to the 
Restricted level) have basic defense measures against all attacks. As we move up the hierarchy to 
the Secret and Top Secret levels, we have more layers of security. 
Next, we demonstrate our defense strategies against access control and boot-stage attacks.

\subsection{Defense against Access Control Attacks}
\begin{figure}[h!]
\centering
\includegraphics[width=1.02\linewidth,scale=1]{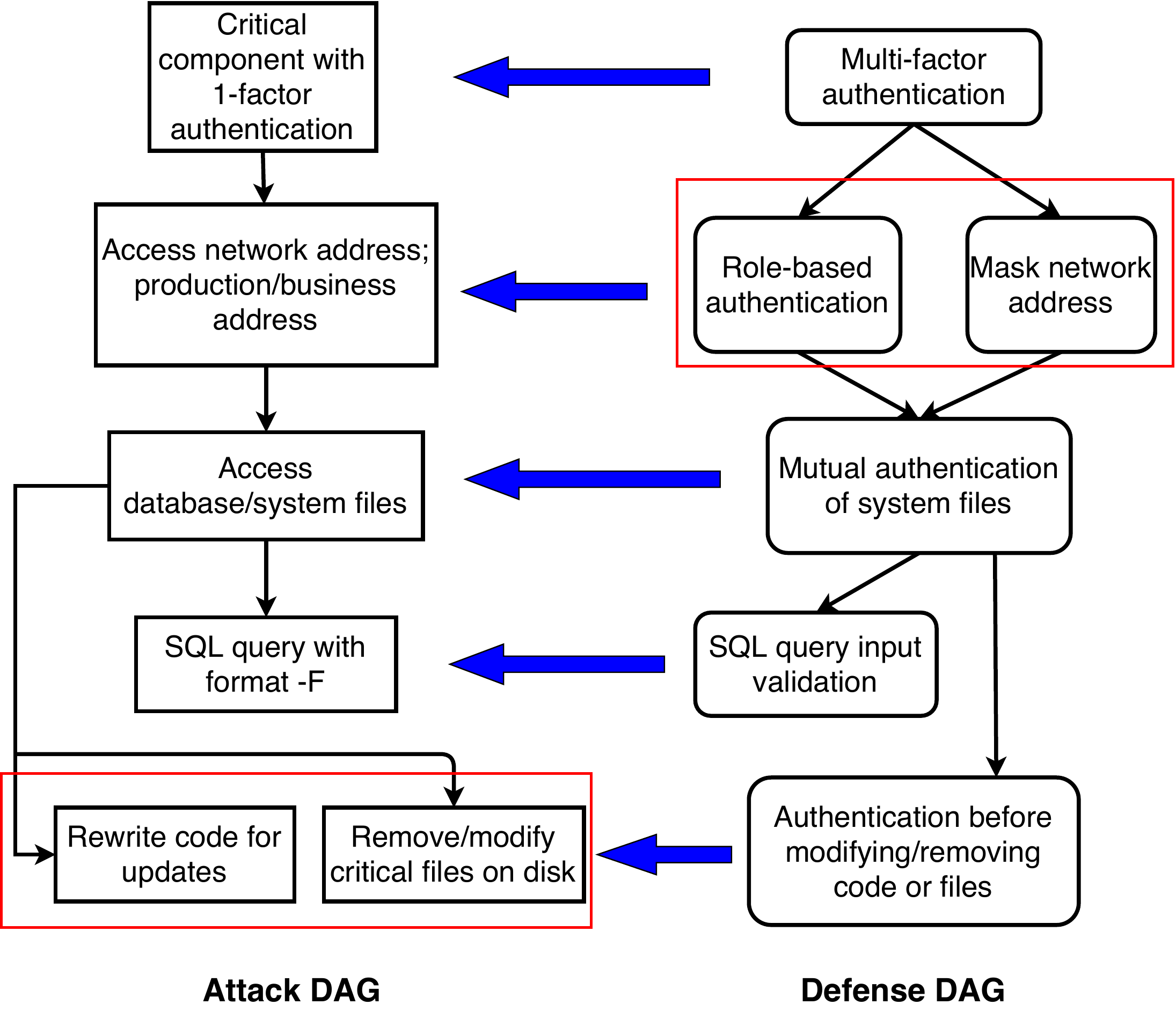}
\caption{Defense at the \textbf{Top Secret} level against access control and privilege escalation 
exploits. The DAG on the left depicts the attack DAG and the DAG on the right depicts the defense 
DAG. The arrows indicate the basic blocks of the defense DAG making the corresponding basic 
blocks of the attack DAG non-operational.} 
\label{Def_ac_ts}
\end{figure}

Access control and privilege escalation attacks are the most common
amongst real-world CPS/IoT attacks, as shown in
Fig.~\ref{fig:histo_attacks}. Access control attacks involve an
unauthorized entity gaining access to a classified resource, thus
compromising its confidentiality and/or integrity. Privilege escalation
attacks involve an entity exploiting a vulnerability to gain elevated
access to resources that it is not permitted to access. Implementation
of strict policies can protect against such attacks. These security
policies include multi-factor authentication, access control lists, and
role-based access control. More layers
of authentication, authorization, and network masking can be added for more sensitive resources. An example of a defense DAG is shown in 
Fig.~\ref{Def_ac_ts}.

In Fig.~\ref{Def_ac_ts}, we demonstrate the security measures deployed in a resource that is
classified as Top Secret. A Top Secret resource is generally a computing device with more
computing resources, the exploit of which can lead to a high-impact damage like the compromise of 
data confidentiality and integrity in a Top Secret data storage server. Hence, we should have 
multiple layers of authentication and network address masking, as shown in Fig.~\ref{Def_ac_ts}. 
In a resource classified as Restricted, there can be only one layer of authentication. More layers 
of authentication may be added for a Secret-level resource depending on the resources it has.

\subsection{Defense against Boot-stage Attacks}
This category of attacks is the most complicated among all the categories. While other attacks can be 
launched at the application level, these attacks typically require root access and have to be launched at the system level.

\begin{figure}[h!]
\centering
\includegraphics[width=1\linewidth,scale=1]{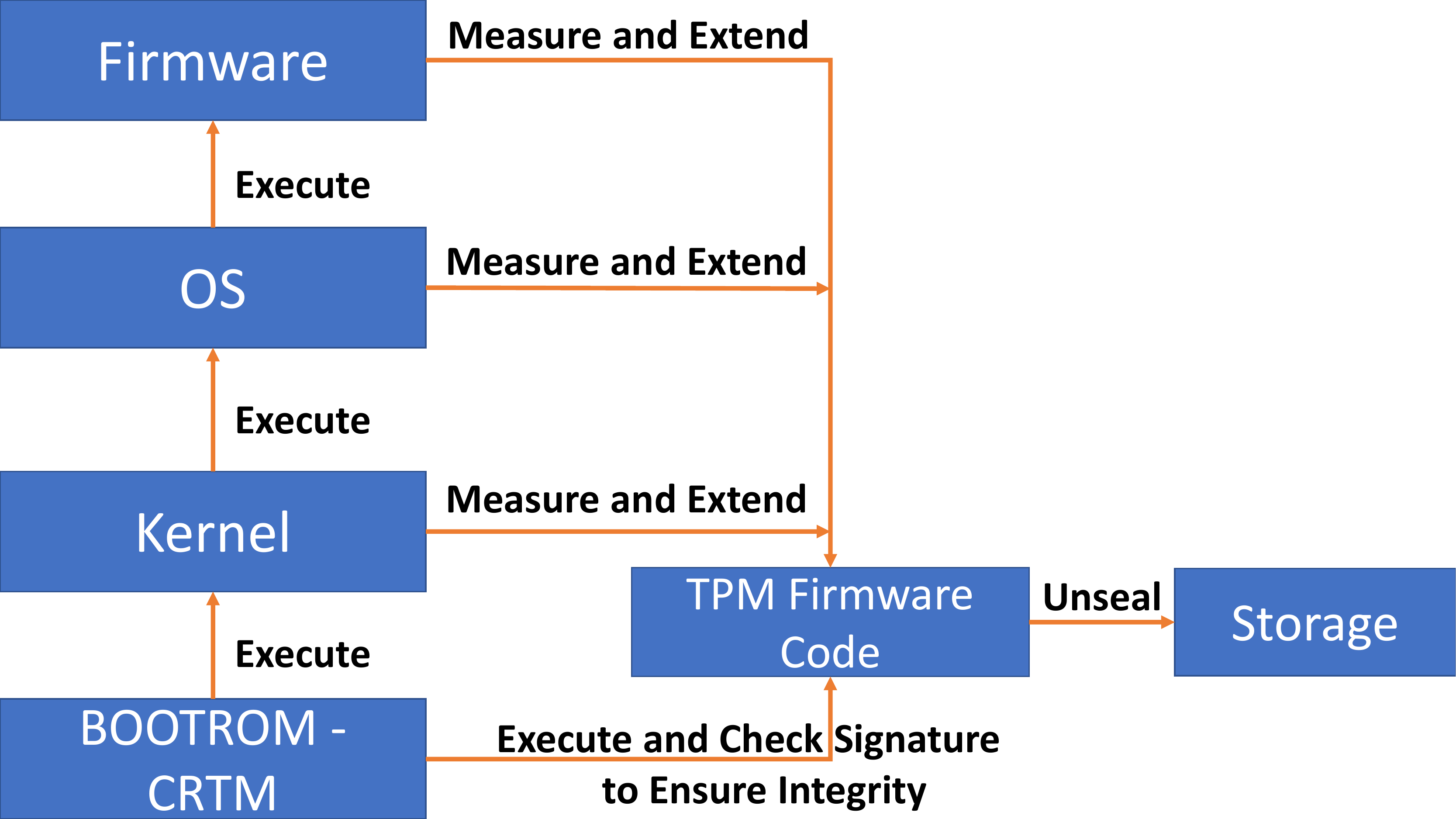}
\caption{Defensive measures against Boot-stage attacks} 
\label{fig:Def_boot}
\end{figure}

To defend against such attacks, a core root of trust for measurement is required along with a trusted platform module (TPM) or a hardware security module. These are generally present at a level lower than the kernel and sometimes referred to as the trusted computing base (TCB). In Fig.~\ref{fig:Def_boot}, the BOOTROM serves as the TCB. Defense against boot-stage attacks involves a series of hierarchical and chained hash checks of binary files and secret keys stored in the Platform Configuration Register (PCR) of the TPM. The PCR is inaccessible to all entities except the TPM. The detection of an incorrect hash value at any stage of the boot sequence causes the boot sequence to halt due to the detection of an illegal modification of the binary boot files and/or the secret(s). Fig.~\ref{fig:Def_boot} gives an overview of the hash checks and execution of binary files at 
various levels.

\section{Conclusion}
\label{section:Conclusion}
The rapid advancement of CPS/IoT-enabling technologies, like 5G communication 
systems and ML, increases the scope of their applications manifold.  Unfortunately, this also increases the attack surface of such systems that can often result in catastrophic effects. We have demonstrated how ML can be used at the system and network levels to detect possible vulnerability exploits across the hardware, software, and network stacks of CPS/IoT. We discovered 10 unexploited attack vectors and 122 novel exploits using the proposed methodology and suggested appropriate defense measures to implement a tiered-security mechanism. We hope that this methodology will prove to be helpful for proactive threat detection and incident response in different types of CPS/IoT frameworks.

\section*{Acknowledgments}
This work was supported by NSF under Grant No. CNS-1617628.

\bibliographystyle{IEEEtranN}
\bibliography{bibtex/bib/IEEEexample}

\begin{IEEEbiography}[{\includegraphics[width=1in,height=1.25in,clip,keepaspectratio]{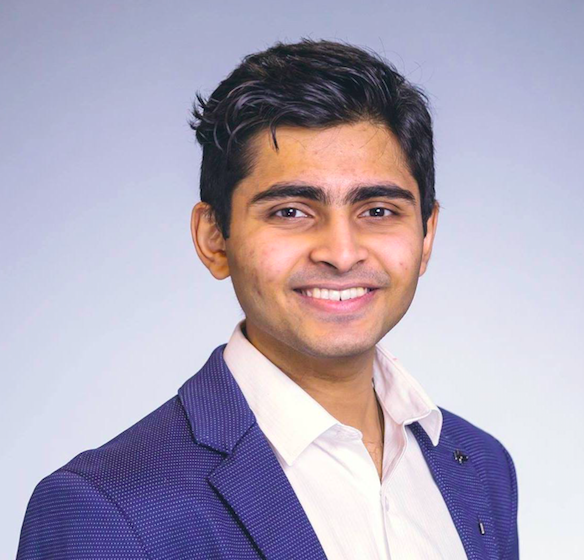}}]{Tanujay Saha}
Tanujay Saha is currently pursuing his Ph.D. degree at Princeton University, NJ, USA. He received his Master's Degree in Electrical Engineering from Princeton University and Bachelors in Technology in Electronics and Electrical Communications Engineering from Indian Institute of Technology, Kharagpur, India in 2017. He has held research positions in various organizations and institutes like Intel Corp., KU Leuven, and Indian Statistical Institute. His research interests lie at the intersection of IoT, cybersecurity, machine learning, embedded systems, and cryptography.
\end{IEEEbiography}

\begin{IEEEbiography}[{\includegraphics[width=1in,height=1.25in,clip,keepaspectratio]{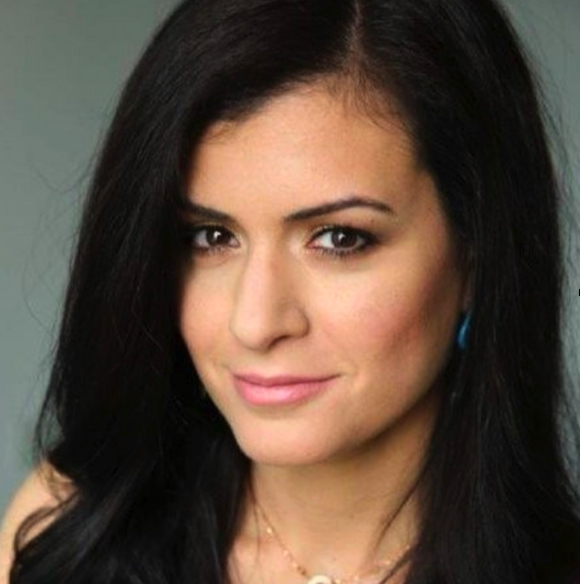}}]{Najwa Aaraj}
Najwa Aaraj is a Chief Research Officer at the UAE Technology
Innovation Institute. She holds a Ph.D. in Electrical Engineering from
Princeton University and a Bachelor’s in Computer and Communications
Engineering from the American University in Beirut. Her expertise lies
in applied cryptography, trusted platforms, secure embedded systems,
software exploit detection/prevention, and biometrics. She has over 15
years of experience working in the United States, Australia, Middle
East, Africa, and Asia with global firms. She has two patents and 15
academic publications. She has worked in a cybersecurity start-up
(DarkMatter). Prior to joining DarkMatter, she worked at Booz \& Company,
where she led consulting engagements in the communication and technology
industry for clients across four continents. She has also held research
positions at IBM T. J. Watson Center, New York, Intel Security Research
Group, Portland, Oregon, and NEC Laboratories, Princeton, New Jersey.
\end{IEEEbiography}

\begin{IEEEbiography}[{\includegraphics[width=1in,height=1.25in,clip,keepaspectratio]{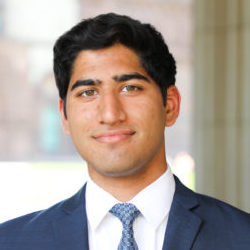}}]{Neel Ajjarapu}
Neel Ajjarapu is currently pursuing his B.S.E in the Department of Electrical Engineering at Princeton University, with a concentration in security and privacy as well as a certificate in technology and society from the Center for Information Technology Policy and Keller Center for Entrepreneurship. He has held intern positions at Microsoft Corp. and One Million Metrics Corp. (Kinetic) in product management and hardware engineering. His current research interests focus on automotive security, embedded systems, and cybersecurity.
\end{IEEEbiography}


\begin{IEEEbiography}[{\includegraphics[width=1in,height=1.25in,clip,keepaspectratio]{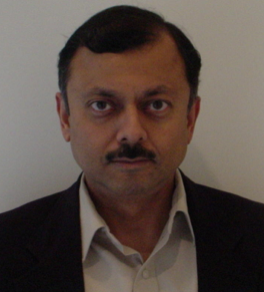}}]{Niraj K. Jha}
Niraj K. Jha received the B.Tech. degree in electronics and electrical 
communication engineering from I.I.T., Kharagpur, India, in 1981, and the 
Ph.D. degree in electrical engineering from the University of Illinois at
Urbana-Champaign, Illinois, in 1985. He has been a faculty member of the
Department of Electrical Engineering, Princeton University, since 1987.
He was given the Distinguished Alumnus Award by I.I.T., Kharagpur. He
has also received the Princeton Graduate Mentoring Award. He has served
as the editor-in-chief of the IEEE Transactions on VLSI Systems and as
an associate editor of several other journals. He has co-authored five
books that are widely used. His research has won 20 best paper awards or
nominations. His research interests include smart healthcare,
cybersecurity, machine learning, and monolithic 3D IC design. He has
given several keynote speeches in the area of nanoelectronic design/test
and smart healthcare. He is a fellow of the IEEE and ACM.
\end{IEEEbiography}




\end{document}